\documentclass[pdflatex,sn-basic]{sn-jnl}

\usepackage{graphicx}%
\usepackage{multirow}%
\usepackage{amsmath,amssymb,amsfonts}%
\usepackage{amsthm}%
\usepackage{mathrsfs}%
\usepackage[title]{appendix}%
\usepackage{xcolor}%
\usepackage{textcomp}%
\usepackage{manyfoot}%
\usepackage{booktabs}%
\usepackage{algorithm}%
\usepackage{algorithmicx}%
\usepackage{algpseudocode}%
\usepackage{listings}%
\usepackage{xcolor}

\theoremstyle{thmstyleone}%
\theoremstyle{thmstyletwo}%

\theoremstyle{thmstylethree}%

\begin{document}

\title[Article Title]{The r-Process: History, Required Conditions, Astrophysical Sites, and Observations}


\author*[1,2]{\fnm{Friedrich-K.} \sur{Thielemann}}\email{f-k.thielemann@unibas.ch, ORCID 0000-0002-7256-9330}

\author[3]{\fnm{John J.} \sur{Cowan}}\email{jjcowan1@ou.edu, ORCID 0000-0002-6779-3813}
\equalcont{These authors contributed equally to this work.}


\affil*[1]{\orgdiv{Department of Physics}, \orgname{University of Basel}, \orgaddress{\street{Klingelbergstrasse 82}, \city{4056 Basel}, \country{Switzerland}}}

\affil[2]{\orgdiv{Theory Group}, \orgname{GSI Helmholtz Center for Heavy Ion Research}, \orgaddress{\street{Planckstrasse 1}, \city{64291 Darmstadt}, \country{Germany}}}

\affil[3]{\orgdiv{Homer L. Dodge Department of Physics \& Astronomy}, \orgname{University of Oklahoma}, \orgaddress{\street{440 W. Brooks St.}, \city{Norman}, \postcode{73019}, \state{Oklahoma}, \country{USA}}}


\abstract{This review of the rapid-neutron-capture (i.e. r-) process starts with determining the Solar System r-abundance pattern via first obtaining (and subtracting) the contribution from the slow-neutron capture (s-) process. We emphasize the extensive work in this area by our late colleague Roberto Gallino and continue in an overview, concentrating on attempts to reproduce the solar r-process pattern with historical site-independent approaches, based on nuclear physics far from stability. In a second step we address the existing proposals for astrophysical sites. Among stellar observations we start with available observations of individual events before analyzing low-metallicity stars, which witness r-process contributions in the early Galaxy. We conclude with a comparison of observations and model predictions, focusing on our present ability to identify the responsible individual astrophysical sites by their imprint in Galactic evolution.}




\maketitle

\section{Introduction}\label{sec1}
\begin{figure}[h!]
\centering
\includegraphics[width=0.8\textwidth]{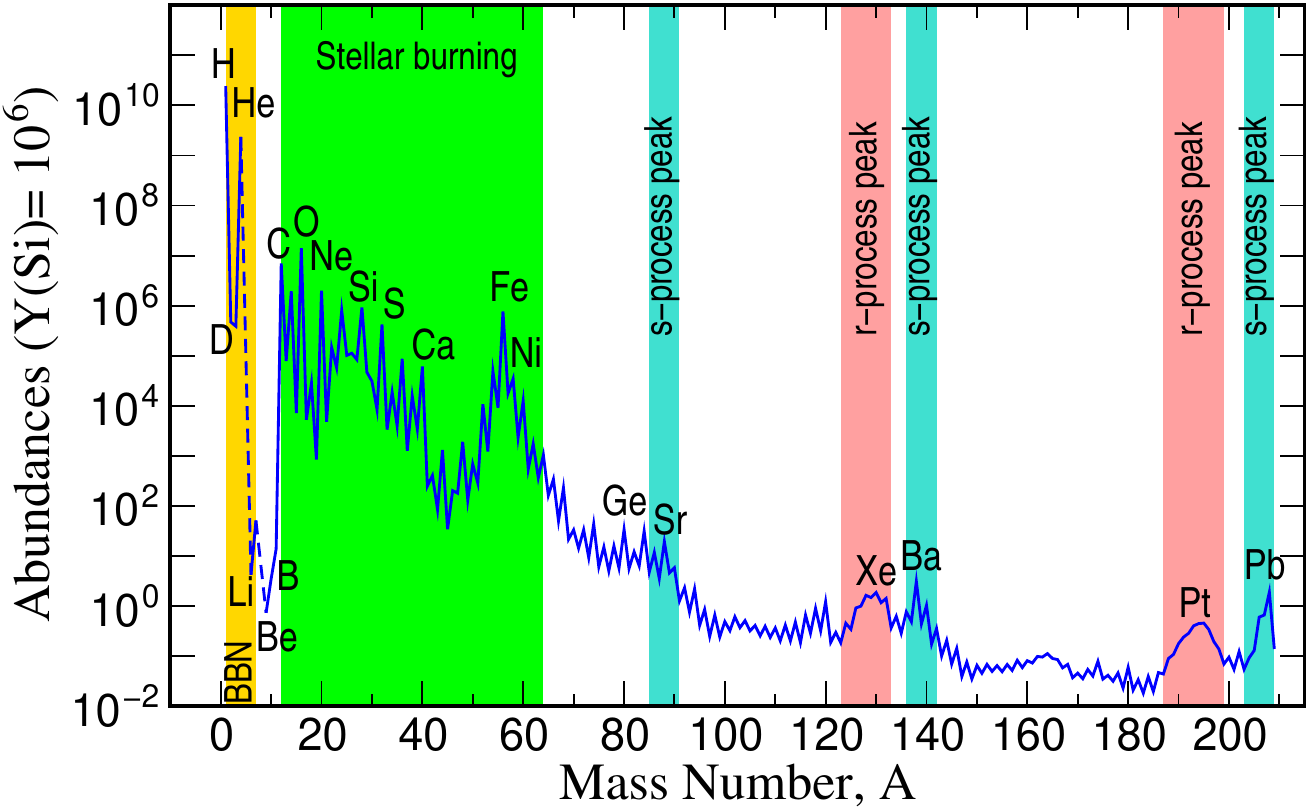}
\caption{Solar System abundances $Y$, scaled to $Y$(Si)=10$^6$ from \cite{Cowan.Sneden.ea:2021}. Abundances are compiled from meteorite measurements \citep{Lodders19,Lodders:2021} and solar spectra \citep{Asplund.Grevesse.ea:2009,Asplund.Amarsi.ea:2021} and are indicative of the values at time of formation of the Solar System; reprinted with permission from \cite{Cowan.Sneden.ea:2021}.}\label{fig1}
\end{figure}

First attempts to provide a complete set of elemental abundances in the solar system were undertaken by \cite{Suess.Urey:1956}. Among other aspects, these inspired \cite{Burbidge.Burbidge.ea:1957} and \cite{Cameron:1957} to lay out the groundwork for understanding the origin of the elements in the Universe via the contributions from stellar evolution and explosions. Further improvements in determining solar system abundances have been contributed by \cite{Cameron:1973}, \cite{Anders.Grevesse:1989}, \cite{Asplund.Grevesse.ea:2009}, \cite{Asplund.Amarsi.ea:2021}, and finally \cite{Lodders:2021} and \cite{Lodders.Bergemann.Palme:2025}, see Fig.\ref{fig1}. 

These ground-breaking explorations made clear that isotopes and elements heavier than iron can only be made by neutron capture due to the rising electrostatic repulsion with increasing nuclear charge. Solar abundances indicated that two different neutron-capture processes dominate, either the
slow-neutron-capture process, i.e. the s-process, or the rapid-neutron-capture
process, i.e. the r-process  - in roughly equal amounts. In order to determine the r-process abundance pattern an essential prerequisite was the understanding of s-process abundances, where Roberto Gallino (to whom the present volume is dedicated) played an essential role. Improving knowledge of the s-process abundances (see also the next Section), and subtracting them from the overall solar abundance pattern, resulted with continuously enhanced accuracy in the decomposition into s-process and r-process components  \citep[see e.g.][]{Cowan.Thielemann.Truran:1991,arlandini99,Goriely:1999,Sneden.Cowan.Gallino:2008,Prantzos.Abia.ea:2020}.
This was achieved via ongoing improvements in understanding of the s-process, that were (besides theoretical modeling) based strongly on the measurements of neutron capture cross sections for nuclei close to stability, which are part of the s-process path. Two groups should be mentioned here, the Oak Ridge group \citep{Macklin.Gibbins:1965} and the Karlsruhe group \citep{Kaeppeler.Beer.Wisshak:1989}, permitting major progress by a collaboration of experimentalists and theorists \citep[see e.g.][]{Kaeppeler.Gallino.ea:2011}. Recently the CERN n\_ToF facility has taken over a large responsibility for these measurements \citep{Domingo-Pardo.NTOF:2023,Milazzo.ea:2025,Patronis.ea:2025}, the first reference still including Franz K\"appeler as coauthor, who jointly with Roberto Gallino contributed immensely to the understanding of the s-process.
Besides the dominant s-process and r-process contributions to the heavy elements,
smaller contributions originate from rarer nucleosynthesis mechanisms, including the p-process
\citep{Burbidge.Burbidge.ea:1957,Cameron:1957, Woosley.Howard:1978,Rayet.ea:1990,Arnould.Goriely:2003,Dillmann.ea:2008,Xiong.Martinez.ea:2024}, the i-process \citep{cowan77,denissenkov17,Denissenkov.ea:2021} and possibly the
LEPP \citep{Travaglio.etal:2004,Cristallo.ea:2015}.

\begin{figure}[h]
\centering
\includegraphics[width=0.75\textwidth]{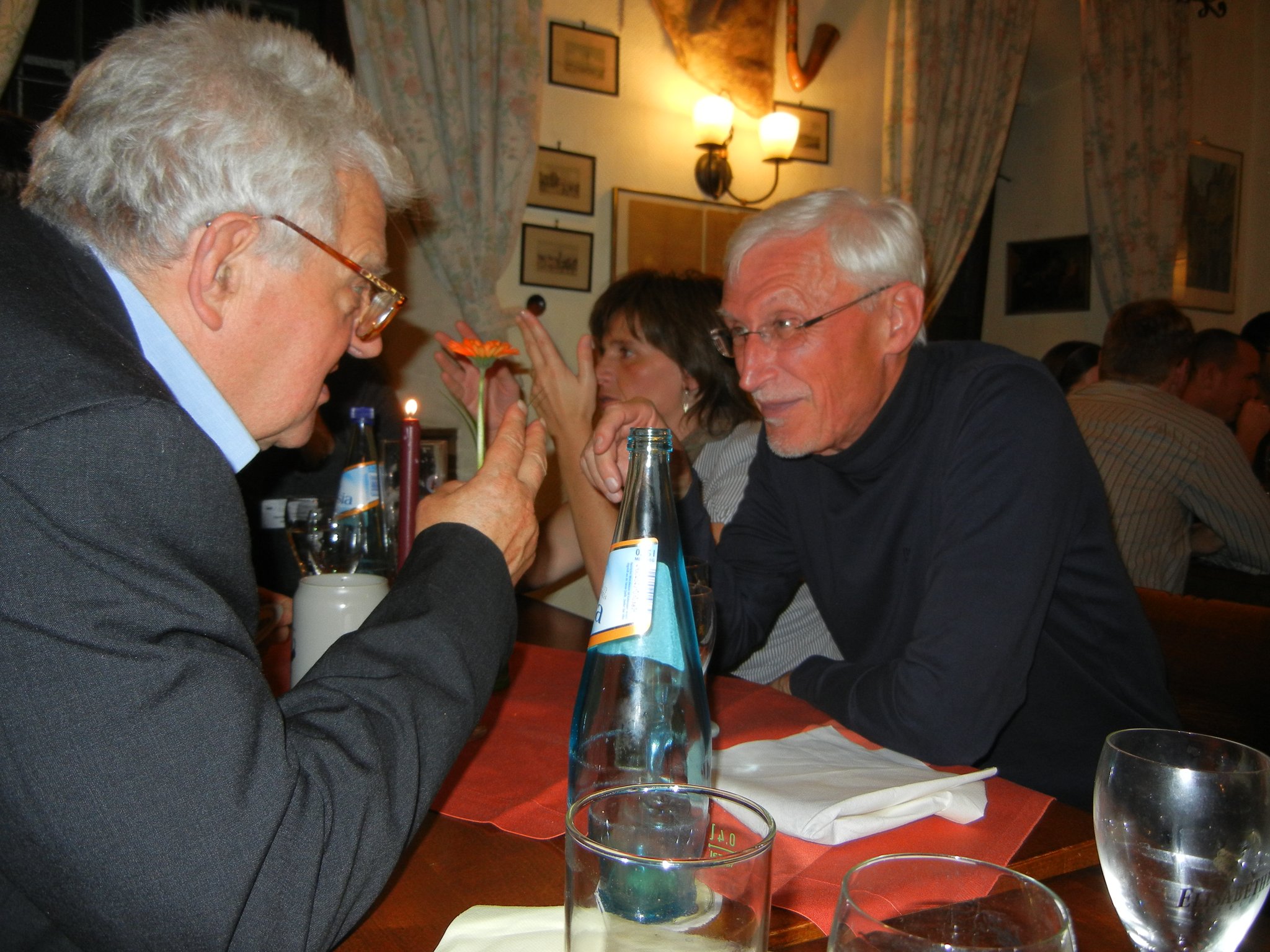}
\caption{Roberto Gallino (left, to whom this volume is dedicated) in animated conversation with Franz K\"appeler (right) during a scientific meeting in Basel, Switzerland in October 2011}\label{fig:robertofranz}
\end{figure}
This paper is intended to be a review of our current understanding of
the r-process, and, at the same time, to honor and
highlight the contributions of our late friend and colleague Roberto Gallino (see Fig.\ref{fig:robertofranz}). Without his contributions towards the understanding of the s-process, a prerequisite for determining the solar system r-process abundances, a detailed examination of the r-process would not have been possible. Therefore, we  first highlight the s-process abundance work necessary to obtain
solar system r-process abundances.
This is followed by a survey over theoretical modeling approaches for possible r-process sites. Finally we will discuss recent r-process abundance observations pointing to such sites as well as their imprint on abundance patterns during the (chemical) evolution of galaxies.


\section{Solar Abundances, the s-Process, and r-Process Residuals}\label{sec2}

\begin{figure}[h!]
\centering
\includegraphics[width=0.65\textwidth]{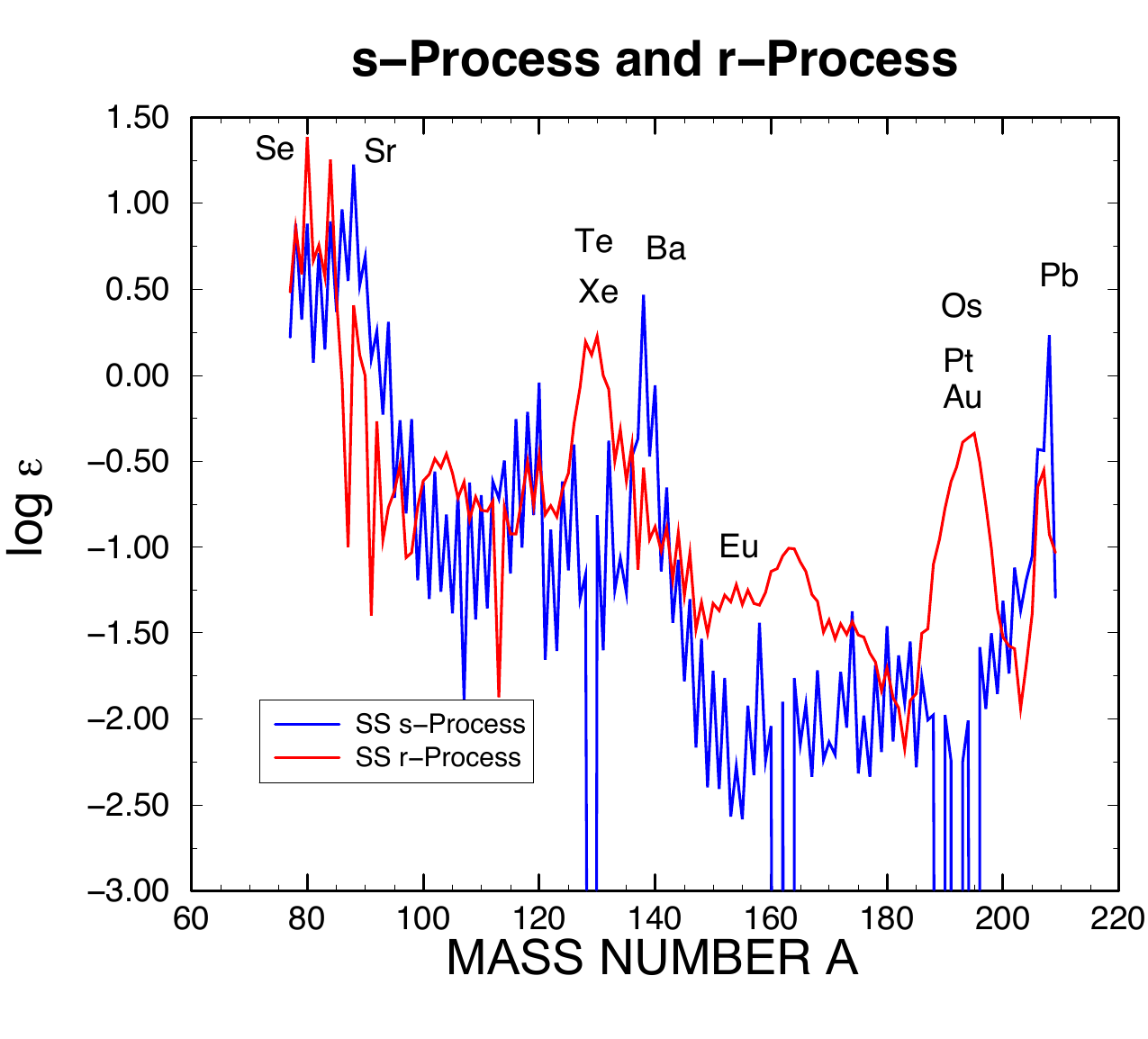}
\caption{A breakdown of the heavy neutron-capture elements into individual s- and r-process contributions. Scale based upon log $\epsilon$ = 12 for the element H; reprinted with permissions from \cite{Cowan.Thielemann:2004}.}\label{fig3}
\end{figure}

Solar system abundance determinations rely heavily on meteoritic measurements and the analysis of the solar spectrum, presently based upon the work of \cite{Lodders19,Lodders:2021,Lodders.Bergemann.Palme:2025} and \cite{Asplund.Grevesse.ea:2009,Asplund.Amarsi.ea:2021} shown in Fig.\ref{fig1}. 
These abundances have resulted from a variety of nucleosynthetic processes integrated over galactic evolution before formation of the solar system.
As mentioned above, it has been known since \cite{Burbidge.Burbidge.ea:1957} and \cite{Cameron:1957} that the majority of all isotopes and elements heavier than iron are synthesized either by the s-process or the r-processes in roughly equal amounts. 
We show in Fig.\ref{fig3} a breakdown of the individual s- and r-process contributions to the total solar system abundances, where the scale is $log\ \epsilon$(H) = 12. This decomposition has only been possible after having quality s-process abundance determinations. 
It is obvious that there are differences in the abundance peaks for the solar system s- and r-process, such that the peaks are offset from each other, and there are defining elements for each nucleosynthetic process, e.g., Os, Pt and Au for the r-process and Pb and Ba for the s-process.

Since the s-process isotopes are stable or formed close to the neutron-rich side of the ``valley of beta stability'', many of their properties can be measured experimentally. With the discovery of Tc (without any stable isotope and $\tau_{1/2}(^{99}$Tc)=$2.1\times 10^5$y) in the envelopes of red giant AGB stars \citep{merrill52}, it was clear that Tc could not have been inherited from previous stellar generations, but that it was formed during stellar evolution of such low and intermediate mass stars. In particular evolved AGB stars have been identified as places where H can be mixed convectively into carbon-rich regions near the He-burning shell, causing the production of $^{13}$C by proton capture on $^{12}$C (and the beta decay of $^{13}$N), which acts as a neutron source via the 
$^{13}$C$(\alpha,n$) reaction. As the s-process timescale is characterized by neutron captures, with (in comparison) shorter beta decays, it can be understood (except for rare branchings between neutron capture and beta decay with comparable timescales) just as a sequence of neutron capture events. Thus, for each mass number $A$ it is characterized by one unique nucleus. \cite{Ulrich:1982} and earlier \cite{Ulrich.Scalo:1972} showed that the neutron production in subsequent He-shell flashes could be represented by neutron exposures
$\tau=\int n_n dt$ (if the $<\sigma v>_{n,\gamma}$ reaction rates for the individual reactions are approximately constant due to the fact that $\sigma\propto 1/v$). 
Thus, the effect of subsequent neutron pulses can be characterized by an integrated neutron exposure, caused in evolved AGB by the $^{13}$C neutron source, which
helps to steadily build up the heavier isotopes and elements. There is one exception, massive stars experience also neutron production in stable He burning via the $^{22}$Ne neutron source, originating from CNO nuclei during He burning. This is the weak s-process producing essentially only nuclei up to $A\approx$80-100, shown first by \cite{Lamb.Howard.ea:1977}. 

The details of the nucleosynthesis, depend critically upon a number of parameters, including the mixing timescales and the characteristics of so-called $^{13}$C
``pockets'' in the envelope of AGB stars. 
After the pioneering model calculations discussed above, \cite{Kaeppeler.Beer.Wisshak:1989} showed nicely how the solar s-process abundances can be reproduced with superpositions of only two sets of neutron exposures $\rho(\tau)$ and their appropriate weights $\rho$: $\rho_1(\tau)\propto \exp(-\tau/\tau_{01})$ and
$\rho_2(\tau)\propto \exp(-\tau/\tau_{02})$. The abundances up to A=80-90 (the weak component with $\tau_{01}$) relate to He burning in massive stars with the $^{22}$Ne neutron source, while the main component (with $\tau_{02}$) describes well the more massive nuclei, which is interpreted
(in terms of an astrophysical setting) by the $^{13}$C source from helium shell flashes with mixed-in hydrogen. 

Based on these early investigations, Roberto Gallino's group followed with a strong impact, extending the s-process studies beyond the nuclear aspects and neutron exposures to realistic  calculations based on stellar evolution, often collaborating with the Karlsruhe group \citep{Busso.ea:1988,Gallino.Busso.ea:1988,arlandini99,busso99,Kaeppeler.Gallino.ea:2011,Bisterzo.ea:2012,Bisterzo.ea:2014,bisterzo15,bisterzo17,Busso.ea:2021}. The resulting excellent understanding of the s-process contributions permitted to also predict high precision solar r-process abundances by subtracting the s-process pattern from solar abundances. This resulted in improved s/r decomposition of solar abundances \citep[see e.g.][and Fig.\ref{fig3}]{Cowan.Thielemann.Truran:1991,arlandini99,Goriely:1999,Sneden.Cowan.Gallino:2008,Prantzos.Abia.ea:2020}. 

\begin{figure}[h!]
\centering
\includegraphics[width=0.7\textwidth]{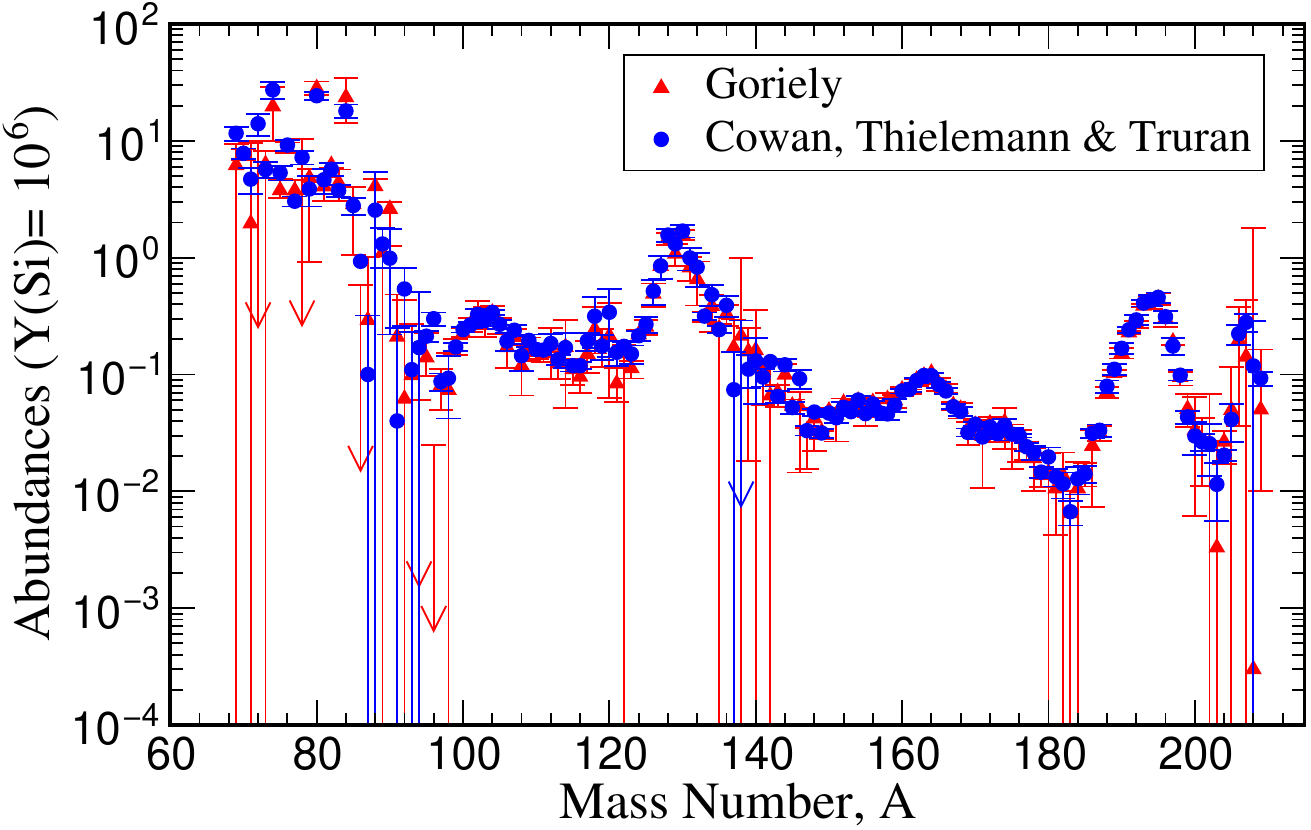}
\caption{Solar r-process abundances (residuals) from \cite{Cowan.Thielemann.Truran:1991} and \cite{Goriely:1999}; reprinted with permission from \cite{Cowan.Sneden.ea:2021}.}\label{fig4}
\end{figure}

\section{r-Process Nucleosynthesis Predictions, Site-Independent and Parametrized Studies}\label{sec3}
The aim of the present review is to understand the origin of the solar system r-process contribution, represented by the difference of the solar and s-process abundances (also called the r-process residuals) and displayed in Fig.\ref{fig4}. Shown are two very similar results obtained by
\cite{Cowan.Thielemann.Truran:1991}
and \cite{Goriely:1999}, based upon the s-process work discussed in the previous section \citep[see updates in][]{Prantzos.Abia.ea:2020}. 
This r-process abundance pattern will be employed to constrain the astrophysical conditions and nuclear constraints under which the r-process operates. While historically the high neutron-to-seed ratios that were required for an r-process to operate and produce heavy elements, including those beyond Pb and Bi, were thought to be linked to explosive environments, a  direct connection to astrophysical sites had to wait for many years. Even today a number of options are still discussed. The present section is meant to explore the necessary conditions for a successful r-process operation, either just (i) by experiencing very high neutron densities in comparison to the s-process, by looking for explosively expanding matter (iii) with different thermodynamic conditions (i.e., entropy), (iii) with different initial abundance patterns ($Y_e$), which after charged-particle freeze from nuclear statistical equilibrium, meet the neutron-density requirements, or finally (iv) whether shocked matter with initial He and e.g. $^{13}$C or $^{22}$Ne abundances can generate the required neutron densities. Thus, these studies are not fully site-independent except for (i), but they explore how r-process conditions can be attained via different environment and thermodynamic options. These can be utilized as a tool, leading eventually to identifying the astrophysical site(s) for r-process nucleosynthesis.

\subsection{Tests with varying neutron densities}
\label{sec:ndens}
In a first approach we would like to focus solely on early attempts which addressed the question how to obtain the solar r-process abundances by studying nucleosynthesis results in a way independent from possible astrophysical sites, just utilizing neutron densities, temperatures, and duration times of the process.
In an early study \cite{seeger.fowler.clayton:1965} assumed a constant temperature and
neutron density. This, together with the knowledge of the neutron capture ($n,\gamma$),
photodisintegration ($\gamma,n$), and beta-decay rates $\lambda_\beta$, permits to define a flow path on the neutron-rich side of
the valley of beta stability. Competition between neutron capture and inverse photodisintegration reactions, for
the specified conditions, determines the distribution of nuclei along each isotope chain, while the longer
beta-decay lifetimes of the involved isotopes dictate the rate of buildup to higher $Z$.
\cite{seeger.fowler.clayton:1965} realized already in these early days, that different conditions are required to obtain good fits to the three r-process peaks around $A=80$ (neutron shell closure $N=50$), $A=130$ ($N=82$), and $A=195$ ($N=126$).
\cite{Cameron.Delano.Truran:1970}
argued that a proper dynamical study of r-process synthesis should be based on the
time-dependent conditions that are expected in the appropriate astrophysical environment. But this relies on the knowledge of astrophysical sites, which we will discuss later.
\cite{Kodama.Takahashi:1975} still followed the same classic approach, as did \cite{Kratz.Bitouzet.ea:1993}, still assuming a set of constant neutron densities $n_n$ and temperatures $T$, but following the time evolution via beta decays along an r-process path. The latter is determined by the so-called $(n,\gamma$)-($\gamma,n)$ equilibrium.
If beta-decays are long in comparison to neutron captures and reverse photodisintegrations, the abundances of neighboring nuclei
$(Z,A)$ and ($Z,A$+1) are determined by production and destruction via neutron capture and via inverse photodisintegration

\begin{equation}
\dot Y(Z,A)=\lambda_{\gamma,n}(Z,A+1)Y(Z,A+1)-<\sigma v>_{n,\gamma}(Z,A)n_nY(Z,A).
\label{eq:1}
\end{equation}
\vskip 1mm

These fast reactions (in comparison to the longer beta decays) lead to an equilibrium in isotopic chains. One can realize this by having a look at the neutron capture reaction rates $<\sigma v>_{n,\gamma}$ in Table 1 of \cite{Cowan.Thielemann.Truran:1991}. Even 10 units off stability typical reaction rates $<\sigma v>_{n,\gamma}$ are of the order $10^{-17}$cm$^3$s$^{-1}$, resulting in  reaction timescales 

\begin{equation}    
\tau_{n,\gamma}= {{1}\over {(n_n<\sigma v>_{n,\gamma}})} 
\end{equation} 

\noindent of $10^{-3}$, $10^{-5}$, and $10^{-7}$s for environments with neutron densities $n_n=10^{20}$, $10^{22}$, $10^{24}$ cm$^{-3}$. The reverse rate, given by detailed balance, can be determined via 
    
\begin{align}
    \lambda_{\gamma,n}(Z,A+1) = &\frac{1}{\tau_{\gamma,n}(Z,A+1)} = \frac{g_n G(Z,A)}{G(Z,A+1)} \left(\frac{A}{A+1}\right)^{3/2}\nonumber\\
    &\times \left(\frac{m_u kT}{2\pi \hbar^2}\right) ^{3/2}\langle \sigma v \rangle_{n,\gamma}(Z,A) \exp\left(-\frac{S_n(Z,A+1)}{kT}\right)
\end{align}
with $g_n=(2\times 1/2 +1)=2$ for neutrons, $G$ being the partition functions of the participating nuclei, and $m_u$ the nuclear mass unit. Utilizing neutron separation energies $S_n$ of 2 to 4 MeV and temperatures $T=1.5\times 10^9$K, the photodisintegration timescale $\tau_{\gamma,n}=1/\lambda_{\gamma,n}$ is of the same order as $\tau_{n,\gamma}$. i.e. $5.4\times 10^{-7}$s for $S_n=2$MeV and $T=1.5\times 10^9$K. Therefore, such an equilibrium for reactions within an isotopic chain is fully justified, i.e., $\dot Y(Z,A)$=0 leads to
an abundance ratio ${Y(Z,A+1)/Y(Z,A)}=[{<\sigma v>_{n,\gamma}(Z,A)/ \lambda_{\gamma,n}(Z,A+1)}]n_n
$ in this ($n,\gamma$)-($\gamma,n$) equilibrium (also called the waiting-point approximation). The corresponding distribution in each isotopic chain (dominated by the maximum abundance in one nucleus or very few nuclei) has to wait to decay via the longer beta decays to the next isotopic chain $Z+1$.
Expressing the reverse photodisintegration via
detailed balance \citep[see above or][]{Cowan.Sneden.ea:2021} and utilizing
the neutron-separation energy of nucleus ($Z,A+1$), $S_n(Z,A+1)$, leads to
abundance ratios dependent only on $n_n$, $T$ and
$S_n$ (when approximating the partition function ratio by 1).

\begin{equation}
{{Y(Z,A+1)}\over {Y(Z,A)}}=n_n {{G(Z,A+1)}\over{2G(Z,A)}}
\Bigl[{{A+1}\over {A}}\Bigr]^{3/2} \Bigl[{{2\pi
\hbar^2}\over {m_ukT}}\Bigr]^{3/2} {\rm exp}(S_n(Z,A+1)/kT). 
\label{eq:2}
\end{equation}

\begin{figure}[h!]
    \centering
    \includegraphics[width=0.65\linewidth]{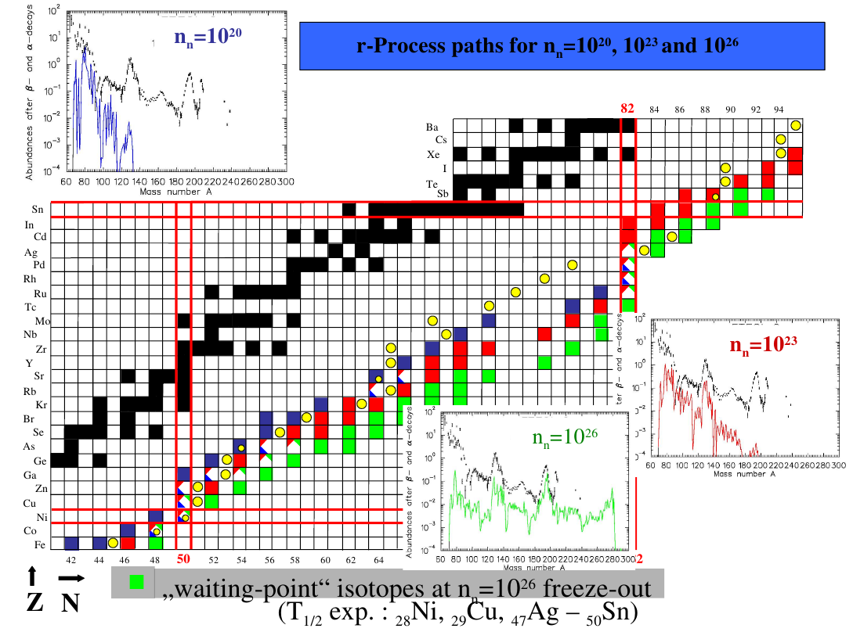}
    \caption{Three r-process paths for $n_n=10^{20}$cm$^{-3}$ (blue), $n_n=10^{23}$cm$^{-3}$ (red), and $n_n=10^{26}$cm$^{-3}$ (green) and the same $T=1.35\times 10^9$K. Inserts show abundances attained after 1.2, 1.6, and 2.5s for the different conditions, in all cases starting with a seed nucleus at $Z=26$. The bottom line indicates for which nuclei experimental beta-decay half lives existed at that time ($^{79}$Cu, $^{80}$Zn, $^{81}$Ga,
$^{91,92}$Br, $^{97-100}$Rb, $^{130}$Cd, $^{131}$In, \citep[see e.g.][]{Gill.Casten.ea:1986,Kratz.Gabelmann.ea:1986}. Due to the different distances of the path from stability (with declining beta-decay half-lives), comparable timescales allow for different endpoints in $A$; courtesy of K.-L. Kratz.}
    \label{fig:rprocesspath}
\end{figure}

This procedure results in maximum abundances in each isotopic chain, which are located at the same $S_n$. Such locations are shown for three different environmental conditions in 
Fig.\ref{fig:rprocesspath}. $S_n$ introduces the dependence on nuclear masses, i.e. a nuclear-mass
model for these very neutron-rich unstable nuclei. The information for Fig.\ref{fig:rprocesspath} has relied mostly on theoretical predictions \citep[initially based on FRDM and ETFSI mass models][]{Aboussir.Pearson.ea:1995, Moeller.Nix.Kratz:1997}, but experiments are progressing (slowly) towards the neutron-drip line, defined by $S_n=0$. 
\begin{figure}[h!]
    \centering
    \includegraphics[width=0.48\linewidth]{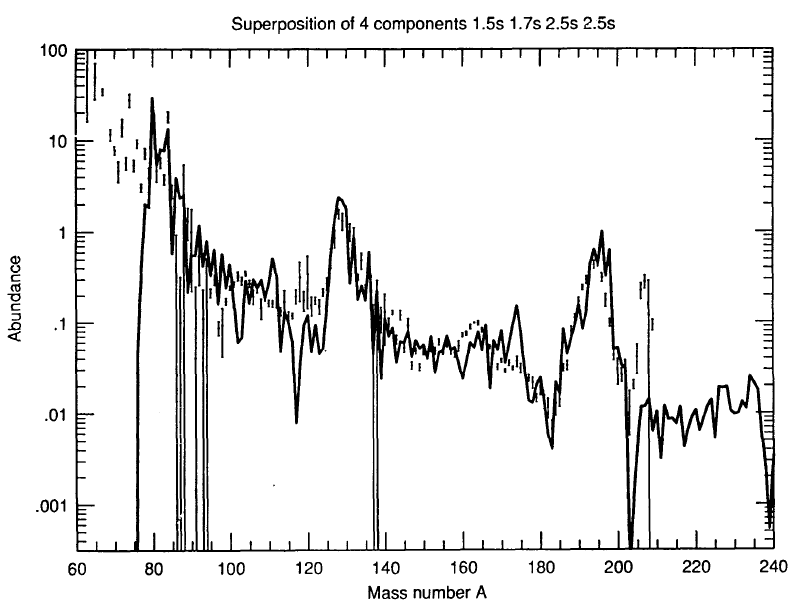}
    \includegraphics[width=0.5\linewidth]{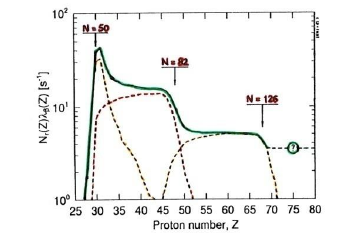}
    \caption{Left panel: result of four superpositions in $n_n$, see text, right panel: indication that a steady flow of beta decays is attained in the r-process between neutron shell closures, underlined by close to constant $Y(Z)\lambda_\beta(Z)$ products, only hindered by long beta decays in the r-process paths in kinks at magic numbers, deviating towards stability at shell closures; adapted from \cite{Freiburghaus.Rembges.ea:1999}.}
    \label{fig:overall}
\end{figure}

Under the assumption of
an ($n,\gamma$)-($\gamma,n$) equilibrium, no
detailed knowledge of neutron-capture cross sections is needed, but the relative neutron separation energies $S_n$, i.e. mass differences enter directly.
Although the early results of Fig.\ref{fig:rprocesspath} \citep[see e.g.][]{Kratz.Thielemann.ea:1988,Thielemann.Bitouzet.ea:1993,Kratz.Bitouzet.ea:1993} were obtained while utilizing ($n,\gamma$)-($\gamma,n$) equilibrium in each isotopic chain, time evolution enters via beta decays between them. The three inserts identify the situation at three different durations for the three paths, with $S_n$ values between 2 and 4 MeV. Already with three superpositions ($n_n= 10^{20}$, $10^{22}$, and $10^{24}$cm$^{-3}$) an overall quite reasonable fit to solar r-process abundances could be obtained (with about a 10:3:1 ratio of the weights for the different components).  Adding a fourth component ($n_n=10^{26}cm^{-3}$) can also produce nuclei up to the actinides. Fig.\ref{fig:overall} (left panel) shows such a superposition of four components. Deviations below the r-process peaks at $A$=130 and 195 (shell closures $N$=82 and 126) could be interpreted as incorrect transitions between deformed and spherical nuclei at closed shells and the related effect on binding energies \citep{Chen.Dobaczewski.ea:1995}. \citep[For a recent overview of such effects see e.g.][]{Li.McLaughlin.Surman:2026}. The right panel indicates a steady beta-decay flow between neutron shell closures. Remember, the s-process, with a process speed being determined by neutron capture timescales and their large values at neutron shell closures due to very small capture cross sections, is characterized by a close to a steady flow of neutron captures in between closed shells. The r-process, with its process speed determined by beta-decay timescales, is characterized by a very similar behavior in terms of a steady flow of beta decays ($\lambda_\beta(Z)Y(Z)=const$), except for long half-lives at kinks of the r-process path closer to stability. 
The right panel of Fig.\ref{fig:overall} shows this product of abundances $Y(Z)=\sum_AY(Z,A)$ in the r-process path with beta-decay rates $\lambda_\beta(Z)=1/Y(Z)\sum_A\lambda_\beta(Z,A)Y(Z,A)$ of the corresponding nuclei. The individual dashed lines show these products for each of three components; the overall line shows the products for a superposition of 13 components with five $n_n$ components spread over each order of magnitude in $n_n$.  

\begin{figure}[h!]
    \centering
    \includegraphics[width=0.65\linewidth]{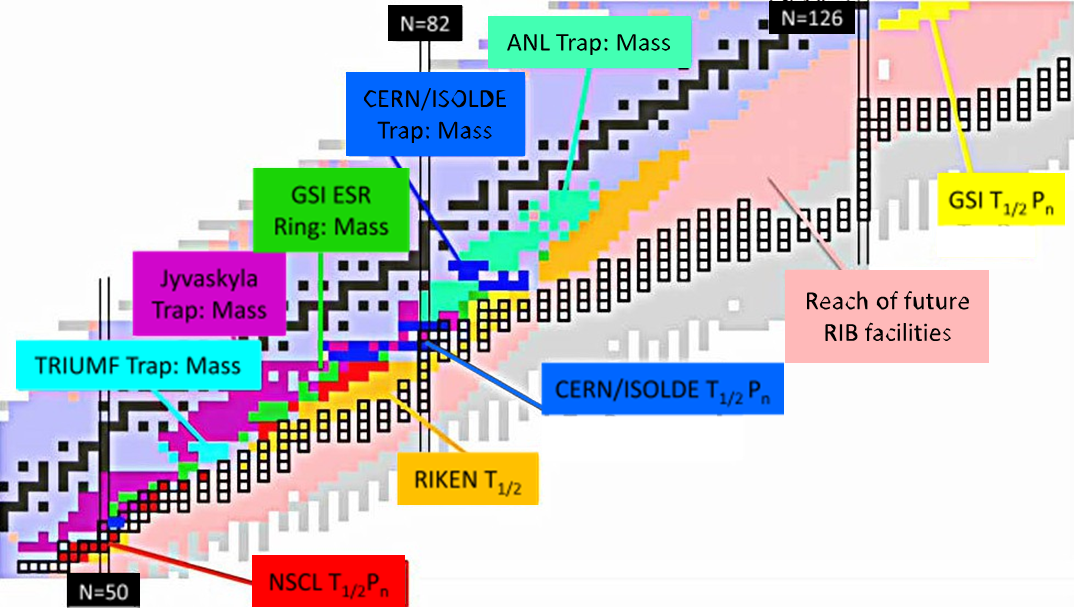}
    \caption{Nuclear chart with stable isotopes marked by black boxes, colored regions (except for pink) indicate experimentally known information (masses, half lives, beta-delayed neutron emission percentages $P_n$). The pink region stands for expected discoveries by world-wide leading and planned rare isotopic beam facilities (RIB), while the grey regions beyond the r-process paths, up to the neutron-drip line, might still be out of reach. The r-process touches isotopes with existing experiments at the N=50 and 82 neutron shell closures (see also Fig.\ref{fig:rprocesspath}); figure reprinted with permission from \citep[][their Fig.21]{Cowan.Sneden.ea:2021}, adapted from \cite{Horowitz.ea:2019}.} 
    \label{fig:paths}
\end{figure}
Fig.\ref{fig:paths} summarizes present experimental efforts on the neutron-rich side off stability. For a detailed discussion of experimental developments for r-process studies see Section IV of \cite{Cowan.Sneden.ea:2021} and e.g. \citep{Horowitz.ea:2019,Podolyak:2023,Ray.Vassh.ea:2024}. Section V of \cite{Cowan.Sneden.ea:2021} includes a review of theoretical nuclear modeling for r-process input, addressing topics of nuclear masses, cross section predictions, beta-decay half lives (and delayed neutron emission) as well as the role of fission. Overviews of recent theoretical efforts and observational indications for fission in low-metallicity stars are given in 
\cite{Petermann.Langanke.ea:2012},
\cite{Goriely.Martinez-Pinedo:2015}, \cite{Giuliani.Martinez.ea:2020},  \cite{Goriely:2023}, \cite{Thielemann.Rauscher:2023},  \cite{Chen.Li.ea:2025}, \cite{Li.McLaughlin.Surman:2026}, and \cite{Roederer.Vassh.ea:2023}. While Fig.\ref{fig:overall} was obtained with the experimental and theoretical input available at the time, the main results stay valid today, also with the gained improvements in nuclear properties far from stability. This includes the conclusion that at least three different sets of conditions are required to fit the r-process abundance pattern over the whole mass range, whether this requires different conditions in the same astrophysical event or different astrophysical events will be discussed in later sections.


\subsection{Entropy Superpositions}
\label{sec:entropies}
\begin{figure}[h!]
    \centering
    \includegraphics[width=0.6\linewidth]{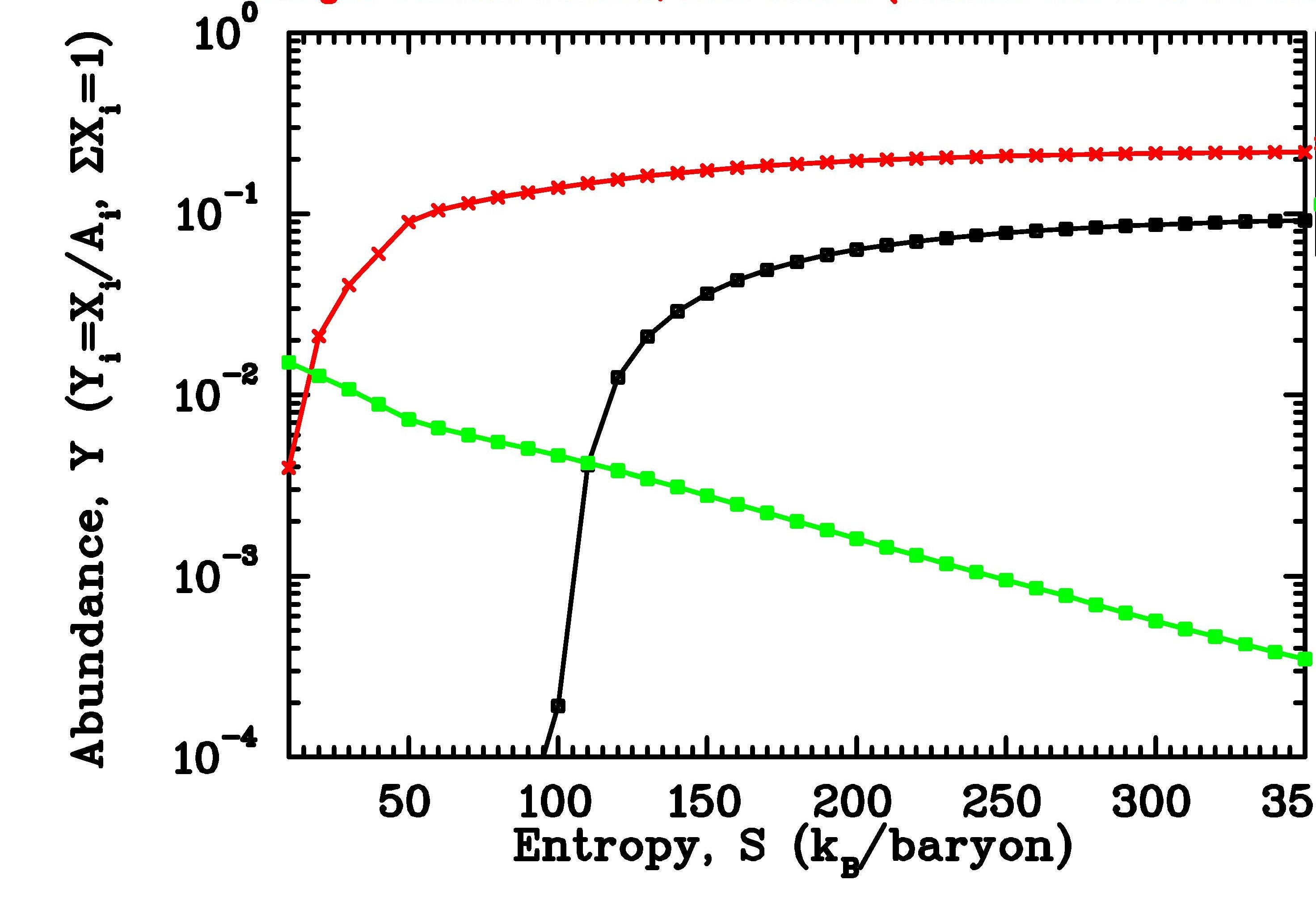}
    \caption{Abundances for $Y_{\alpha}$ (red), $Y_n$ (black), and $Y_{seed}$ (green) resulting after charged-particle freeze-out (before the onset of an r-process) for adiabatic expansions with the entropies shown on the abscissa, $Y_e$=0.45, and an adiabatic expansion with $v_{exp}$=4500km s$^{-1}$; reprinted with permission from \cite{Farouqi.Kratz.ea:2010}, copyright by the AAS.}
    \label{fig:HEW}
\end{figure}

Based on investigations related to high-entropy ejecta from core-collapse supernovae \citep{meyer92,Woosley.Wilson.ea:1994,Takahashi.Witti.Janka:1994,hoffman.woosley.qian:1997} the idea emerged that in the case of high entropies only slightly neutron-rich matter could lead to an r-process. The main reason is that high entropies in explosive silicon burning lead to a highly alpha-rich freeze out of charged-particle reactions. Alpha particles are symmetric with $N=Z$, i.e. if a highly alpha-rich freeze out takes place, 90 to 99 percent - or even more - of the matter can be in alpha particles. Then the small amount of remaining matter, which made it to heavier nuclei, can share the few available neutrons and experience a high neutron-to-seed nuclei ratio, which permits an r-process to take place.
The entropy of a radiation-dominated plasma can be written as 

\begin{equation}
S=(4/3)a(T^3/\rho)[1+(7/4)f(T)] ,    
\end{equation}

where in the second bracket [] the first term stands for pure radiation (i.e. photon domination), while the second term, including $0<f(T)<1$, stands for the contribution of ultrarelativistic electrons and positrons for very high temperatures \citep[see][]{Witti.Janka.Takahashi:1994,Freiburghaus.Rembges.ea:1999}. Rather than assuming constant neutrons densities and temperatures for a given time duration, such calculations have been performed in a fully time-dependent fashion, assuming an initial entropy $S$, an initial temperature $T_0$, a related density $\rho_0$ determined by $S$ and $T_0$, an initial proton/nucleon ratio $Y_e$, and an adiabatic expansion (with adiabatic index $\gamma=4/3$) of a sphere with an expansion velocity
$v_{exp}$, leading to a spherical radius expansion of the type $R(t)=R_0+v_{exp}t$. Starting with a sufficiently high temperature, which allows nuclear statistical equilibrium, and a neutron-to-proton ratio related to the assumed $Y_e$, Fig.\ref{fig:HEW} of \cite{Farouqi.Kratz.ea:2010} shows the resulting abundances $Y$ of alpha particles (red, mass fraction $X_\alpha=4\times Y_\alpha$), neutrons (black), and ``seed nuclei'' in the mass range $A=$50-80 (resulting from transfer of alpha particles, He, to heavier nuclei via the triple alpha or e.g. the $\alpha\alpha n$ reaction).

For $Y_e$=0.5 (symmetric matter) the seed nucleus would be $^{56}$Ni for a normal freeze-out of charged-particle reactions, which could then extend to higher mass numbers for an alpha-rich freeze-out and even further for slightly neutron-rich conditions. For the case of Fig.\ref{fig:HEW} $Y_e=0.45$, $R_0$=130km and $v_{exp}$=4500km s$^{-1}$ were used. It can be seen that for high entropies (despite only a small neutron richness with $Y_e$=0.45) neutron/seed ratios of a few hundred can be attained, which permit a full r-process via neutron captures on the produced seed nuclei. These calculation were performed in a dynamic time-dependent way with varying temperatures and densities and the use of full reaction networks without utilizing an ($n.\gamma$)-($\gamma,n$) equilibrium.

\begin{figure}[h!]
    \centering
    \includegraphics[width=0.8\linewidth]{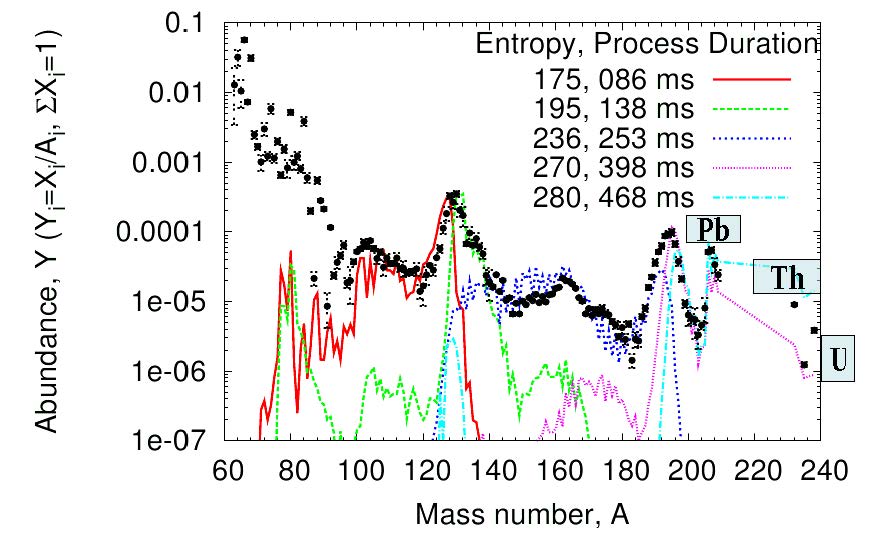}
    \caption{Resulting abundance patterns after r-process processing with the entropies indicated by different colors (for the conditions discussed together with Fig.\ref{fig:HEW}). High entropies and the appropriate highly alpha-rich freeze out conditions are required for producing the heaviest r-process nuclei; reprinted with permission from \cite{Farouqi.Kratz.ea:2010}, copyright by the AAS.}
    \label{fig:entropies}
\end{figure}

The resulting abundance patterns after r-process nucleosynthesis, performed with a full nuclear network and the starting conditions after charged-particle freeze out from Fig.\ref{fig:HEW}, are shown in Fig.\ref{fig:entropies}, labeled with the different entropies utilized. Additional information is provided for each entropy, showing after which time the final abundances shown are attained when a freeze out of available neutrons has occurred. It can be seen that, similar to the simulations with varying neutron densities, varying entropies are required for attaining good fits to the different r-process peaks. Also for these kinds of simulations, a superposition of entropies (and possibly $Y_e$'s) is needed to find a good overall fit to solar r-process abundances. We would like to point also to further studies of such high-entropy environments \citep[see e.g.][]{Kratz.Farouqi.ea:2007,Kratz.Farouqi.ea:2008,Farouqi.Kratz.Pfeifer:2009,Panov.Janka:2009,kratz14}.

\subsection{$Y_e$ Superpositions}
\label{sec:Ye}
As an extension of the previous subsection (entropy superpositions with a relatively high $Y_e$) a more general approach would include variations of all entering parameters, entropy $S$, neutron richness $Y_e$, and the expansion timescale or velocity $\tau_{exp}$ or $v_{exp}$. This was discussed already in \cite{hoffman.woosley.qian:1997} and \cite{Freiburghaus.Rembges.ea:1999}, but here we want to focus on neutron-rich relatively low-entropy conditions which could be of interest in more recently discussed astrophysical environments, like compact binary mergers. 

\begin{figure}[h!]
    \centering
    \includegraphics[width=0.5\linewidth,angle=-90]{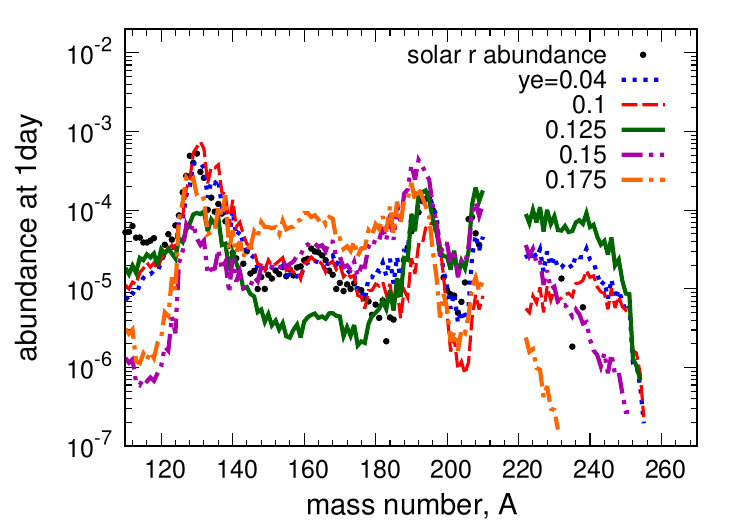}
    \caption{r-process simulations for low entropies of $S\approx$k$_B$/baryon and fast expansions (timescales of the order $5\times10^{-3}$s), utilizing the Duflo-Zuker mass model \cite{Duflo.Zuker:1995} for a range of low $Y_e$ conditions \citep{Thielemann.Wehmeyer.Wu:2020}, see also Fig.\ref{fig:rabund-Aye}. It is seen that large variations in the actinide production are experienced. The highest actinide production is found at Ye = 0.125; reprinted with permission from \cite{Thielemann.Wehmeyer.Wu:2020}.}
    \label{fig:meng-ru}
\end{figure}

Fig.\ref{fig:meng-ru} shows results of expansions with a similar prescription as for the entropy superpositions, but for low entropies of $S\approx5$k$_B$/baryon and short expansion timescales of the order $5\times 10^{-3}$s, i.e. fast expansion velocities from \cite{Thielemann.Wehmeyer.Wu:2020}. 
If the parameter space would have been extended to $Y_e$'s close to 0.4, also abundances in the first r-process peak would have been produced (see also Fig.\ref{fig:rabund-Aye}), and the result is similar to the findings of the previous two subsections: in order to reproduce the solar r-process abundances a superposition of at least three components (= environment conditions) is needed. Here one can see that low $Y_e$'s in the range 0.05 to 0.15 can produce the third peak and the actinides. It can be noticed that small variations in $Y_e$ around 0.125 lead to large variations in the actinide production.

\subsection{Explosive He-burning with neutrons from ($\alpha,n$) reactions}
\label{sec:alphanp}
\begin{figure}[h!]
    \centering
    \includegraphics[width=0.6\linewidth]{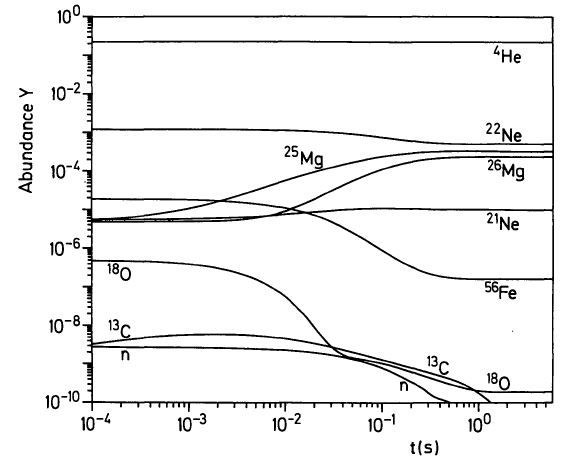}
    \caption{Burning helium shell matter at a temperature of $T=8\times 10^8$K with a density of $\rho=10^4$gcm$^{-1}$ leads over a period of 1s to neutron production of $Y_n=4\times 10^{-9}$ (corresponding to $n_n=2.4\times 10^{19}$cm$^{-3}$. Dominant contributors to the neutron production are $(\alpha,n)$ reactions on $^{22}$Ne, $^{18}$O, and $^{13}$C; reprinted with permission from \cite{Thielemann.Arnould.Hillebrandt:1979}, copyright by ESO.}
    \label{fig:nprod}
\end{figure}

In contrast to the previous subsections we now consider an environment that does not follow the classical or entropy based path, but relies on neutron production via ($\alpha,n$) reactions, similar to the s-process but with higher temperatures and thus higher neutron fluxes. It was first suggested by \cite{Truran.Cowan.Cameron:1978}, \cite{Thielemann.Arnould.Hillebrandt:1979}, and \cite{Cowan.Cameron.Truran:1980}. The initial idea was that in He-burning shells of massive stars the CNO nuclei are transferred to $^{18}$O and $^{22}$Ne, which - when suddenly heated by shock waves - can cause large amounts of neutron production via ($\alpha,n$) reactions.
The results of this neutron production, 
due to the sources $^{22}$Ne, $^{18}$O, and $^{13}$C,
are shown in Fig.\ref{fig:nprod}, taken from \cite{Thielemann.Arnould.Hillebrandt:1979}. 

\begin{figure}[h!]
    \centering
    \includegraphics[width=0.65\linewidth]{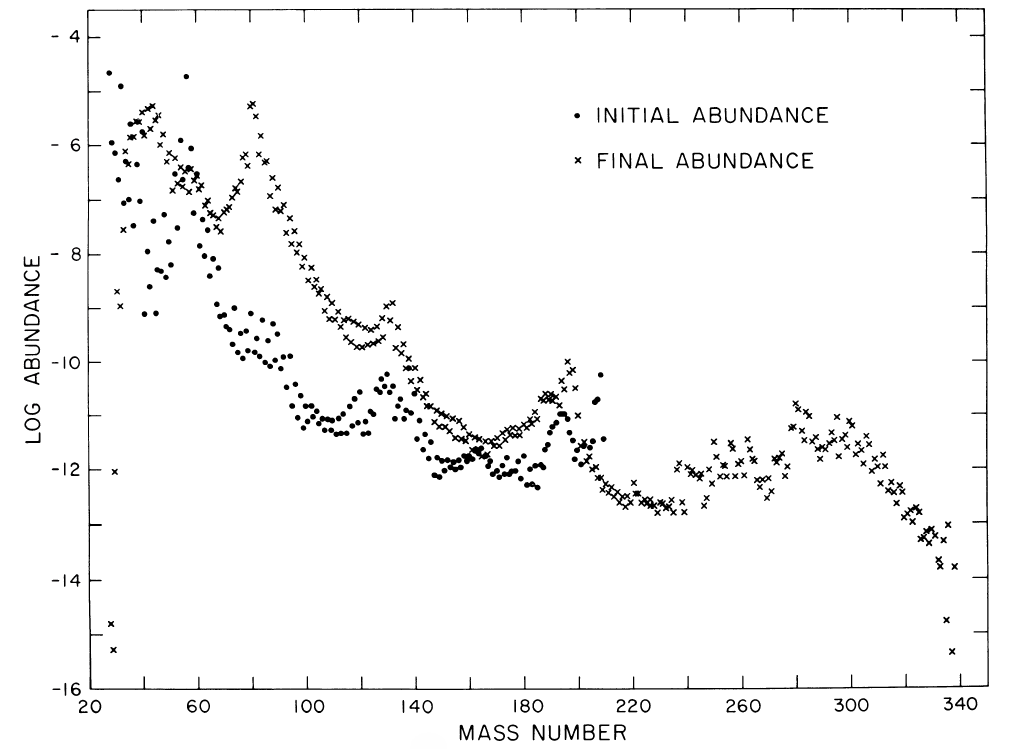}
    \caption{The initial composition (solar system abundances for heavy elements), indicated by black dots, is transferred via neutron captures in an r-process to heavier elements, even beyond the actinides (shown by black crosses before decay; reprinted with permission from \cite{Truran.Cowan.Cameron:1978}, copyright by the AAS.}
    \label{fig:TCC}
\end{figure}

\cite{Truran.Cowan.Cameron:1978} used slightly different conditions with $T=8.5\times 10^8$K and 
$\rho=10^5$gcm$^{-1}$. This led to somewhat higher  neutron densities beyond $10^{20}$cm$^{-3}$. When assuming an initial heavy element solar system composition, the resulting final abundance pattern is shown in Fig.\ref{fig:TCC}. The latter calculations, characterized by relatively low neutron densities for an r-process, were not calculated with the classical $(n,\gamma)$-$(\gamma,n)$ approximation, but instead followed neutron captures by a reaction network. Later investigations by the same authors have utilized enhanced starting abundances of $^{13}$C to attain higher neutron densities \citep{Cowan.Cameron.Truran:1980,Cowan.Cameron.Truran:1983,Cowan.Cameron.Truran:1985}. In \cite{Cameron.ea:1983} it was also shown that in these calculations a steady flow of beta decays is approached, similar to what was shown in Fig.\ref{fig:overall}, but here by using a full nuclear network rather than assuming $(n,\gamma)$-$(\gamma,n)$ equilibrium. This was later discussed in further detail by the same authors \citep{Cameron.Cowan.Truran:1983}.

As a summary of this section (at least subsections \ref{sec:ndens}, \ref{sec:entropies}, and \ref{sec:Ye}), we would like to point out that no single set of conditions can produce the full solar r-process pattern from the first to third peak. Instead, a superposition of at least three conditions or components is required to reproduce the typical r-process pattern as observed in the solar system \citep[see also][]{Kuske.ea:2025}. It will be discussed in the next sections whether those conditions reflect different astrophysical sites (including also variations of section \ref{sec:alphanp}) or whether specific astrophysical sites represent already a superposition of different conditions.
\section{Modelling Astrophysical Sites}
\label{sec:sites}
After having analyzed conditions which support the formation of heavy elements by an r-process in a strategy which utilized  parameter studies, we want to address in this section previously suggested astrophysical sites. In the early past (going back to the 1970s) all suggested scenarios were identified with core-collapse supernovae. But in more recent decades several different classes of core-collapse events have been discussed, which we will review sequentially. Based on early suggestions in the 1970s and 1980s \citep{Lattimer.Schramm:1974,lattimer77,Eichler.Livio.ea:1989}, new scenarios like compact binary mergers were also addressed, starting with more detailed studies in the late 1990s. In this section we attempt to discuss the present status of all these investigations, before focusing in the next section on observations of individual events and their imprints on the chemical evolution of galaxies.

\subsection{Core-collapse Supernovae}
\label{sec:failedsnr}
Before discussing current and viable supernova scenarios, that would result in an r-process, we want to review  some previous historical efforts,  which could not reproduce the necessary r-process requirements when compared to the most up-to-date supernovae simulations \citep[e.g.][]{Burrows.ea:2024,Janka:2025}. 
The initial hopes that the innermost supernova ejecta close to the neutron star remnant would be neutron-rich (similar to the neutron star itself) and lead to a successful r-process site \citep{Hillebrandt.Kodama.Takahashi:1976} have not materialized. It turns out that these ejecta are rather slightly proton-rich due to the neutrino interactions with ejected matter, and the chance that very slightly neutron-rich matter (leading at most to a weak r-process) can be ejected, is still highly questionable \cite[see e.g.][]{Curtis.ea:2019,Ebinger.Curtis.ea:2020,Ghosh.ea:2022}.

In a similar way the suggestions for an r-process in the He shell of massive stars, when the supernova shock wave heats these stellar layers \citep[see Section 3.3 and][]{Truran.Cowan.Cameron:1978,Thielemann.Arnould.Hillebrandt:1979}, are not consistent with our current understanding of stellar models and supernova explosions.  At that time sophisticated stellar evolution models of massive stars were not available, yet, only being put forward shortly thereafter \citep{Weaver.Woosley:1978,Weaver.Woosley:1980}. In addition, the assumed maximum temperatures attained during the passage of the supernova shock wave were smaller than assumed in the proposals by \cite{Truran.Cowan.Cameron:1978} and \cite{Thielemann.Arnould.Hillebrandt:1979}. An efficient r-process was only possible with an enhanced $^{13}$C abundance in the He-shell \citep{Cowan.Cameron.Truran:1980,Cowan.Cameron.Truran:1983,Cowan.Cameron.Truran:1985}, which is not realistic, based on present stellar evolution models. However, these conditions could lead to a weaker n-process, suggested initially by \cite{Blake.Schramm:1976} and found in the Helium shell during the explosion of rotating massive stars by \cite{Choplin.Tominaga.Meyer:2020}.

There exist further suggestions for a successful full r-process in the neutrino wind of core-collapse supernova explosions \citep[e,g,][]{Otsuki.Tagoshi.ea:2000,sumiyoshi01,Thompson.Burrows.Meyer:2001,Terasawa.Sumiyoshi.ea:2001b,Wanajo.Kajino.ea:2001, Terasawa.Langanke.ea:2004,Wanajo:2007}, but in present-day high precision supernova studies these conditions have not yet been materialized. An alternative core-collapse supernova related site has been proposed in ONeMg shells of massive stars \cite{Ning.Qian.Meyer:2007}, but it is probably not a major r-process source.
Following these somewhat negative results for early suggested full r-process sites, we will present in the following subsections currently discussed sites and their predicted abundance features.

\subsubsection{A weak r-process in ``regular'' neutrino-powered supernovae?}
\label{sec:regsne}
Initial core-collapse supernova simulations by \cite{Woosley.Wilson.ea:1994} had resulted in very high entropies in the innermost supernova ejecta, powered by neutrino transport from the hot proto-neutron star, which formed after core collapse -  the so-called high-entropy wind. Follow-up simulations by \cite{Witti.Janka.Takahashi:1994} and \cite{Takahashi.Witti.Janka:1994} could not attain the same high entropies, but showed that - when multiplying their obtained entropies with an appropriate factor - this led to similar r-process results as in \cite{Woosley.Wilson.ea:1994}. However, advancements in neutrino transport \citep{Liebendoerfer.ea:2003,Liebendoerfer.ea:2004,Mezzacappa:2005,Liebendoerfer.Rampp.ea:2005,Liebendoerfer.Whitehouse.Fischer:2009,Fischer.Whitehouse.ea:2009,Fischer.Wu.ea:2020} have led to the important result, that the neutrino interactions do not permit anymore slightly neutron-rich conditions, but rather drive matter to slightly proton-rich conditions via the reactions
\begin{eqnarray}
  \nu_e + n & \rightleftarrows &p+ e^- \\
  \bar\nu_e + p& \rightleftarrows &n+ e^+
\end{eqnarray}
(where the neutron-proton mass difference favors the first reaction for similar neutrino and antineutrino energies).
If the material is subject long enough to these
processes, it reaches an equilibrium between neutrino and antineutrino
captures~\citep{Qian.Woosley:1996,Thompson.Burrows.Meyer:2001,Martinez-Pinedo.Fischer.ea:2017},
resulting in

\begin{equation}
  \label{eq:yeeq}
  Y_e=  Y_{e,\text{eq}} = \left[1+\frac{L_{\bar{\nu}_e}
      W_{\bar{\nu}_e}}{L_{\nu_e}W_{\nu_e}} \frac{\varepsilon_{\bar{\nu}_e} - 2 
      \Delta + \Delta^2/\langle 
      E_{\bar{\nu}_e}\rangle}{\varepsilon_{\nu_e} + 2 \Delta + \Delta^2/\langle
      E_{\nu_e}\rangle}\right]^{-1},
\end{equation}
with $L_{\nu_e}$ and $L_{\bar{\nu}_e}$ being the neutrino and
antineutrino luminosities,
$\varepsilon_{\nu} = \langle E_\nu^2\rangle/\langle E_\nu\rangle$ the
ratio between the second moment of the neutrino spectrum and the
average neutrino energy (similarly for antineutrinos),
$\Delta = 1.2933$~MeV the neutron-proton mass difference, and
$W_{\nu} \approx 1+1.01 \langle E_\nu \rangle/(m_u c^2)$,
$W_{\bar{\nu}} \approx 1-7.22 \langle E_{\bar{\nu}} \rangle/(m_u c^2)$
the weak-magnetism correction to the cross sections for neutrino and
antineutrino absorption~\citep{Horowitz:2002} with $m_u$ being the
nucleon mass.

\begin{figure}[h!]
    \centering
    \includegraphics[width=0.75\linewidth]{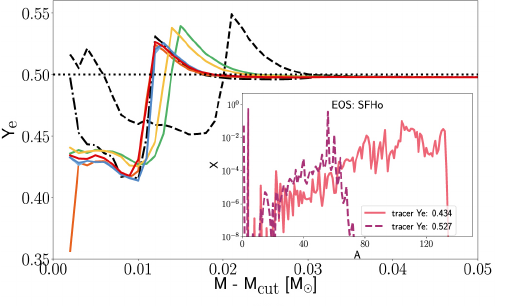}
    \caption{$Y_e$ as a function of the radial mass coordinate in spherically-symmetric PUSH simulations, utilizing several different equations of state for the collapse to neutron star density \citep{Ghosh.ea:2022}. In all cases a similar feature can be viewed: A $Y_e>$0.5 due to neutrino interactions with ejected matter, leading to a $\nu$p-process, and a $Y_e<$0.5 in the very innermost ejecta, stemming from electron capture during collapse to high densities. Whether the matter with $Y_e<$0.5 can be ejected is still uncertain and will need to be verified by high-resolution 3D models. It would lead to a very weak r-process, not going beyond $A=$130 (see inserts for $\nu$p-process and very weak r-process abundance patterns); copyright by the authors.}
    \label{fig:ghosh}
\end{figure}

This led to the postulation of a $\nu$p-process by \cite{Froehlich.Hauser.ea:2006,Froehlich.Martinez-Pinedo.ea:2006} rather than an r-process in the ejected innermost layers. This behavior can be seen in Fig.\ref{fig:ghosh}, obtained with the spherically symmetric PUSH approach \cite{Ebinger.Curtis.ea:2020}. 
Whether the innermost low $Y_e$ part can actually be ejected has to be tested in high resolution 3D simulations \citep{Burrows.ea:2024,Wang.BurrowsG:2024,Janka:2025}, but if ejected it would lead to a very weak r-process up to at most the $A=$130 peak (see insert in Fig.\ref{fig:ghosh}). \cite{Psaltis.ea:2024} studied the combination of a $\nu$p-process and a weak r-process in the inner core-collapse supernova ejecta.

\subsubsection{Magneto-rotational supernovae}
\label{sec:MHD}
More than 10\% of core-collapse supernovae leave magnetars as central remnants \citep{Beniamini.Hotokezaka.Horst.ea:2019,Pardo.Rea.ea:2026}, i.e., neutron stars with magnetic fields of the order $10^{15}$G and pulsar periods of 1-10s, \citep[see e.g. Fig.1 in][]{Borghese.Coti:2026}, which have slowed down since their formation due to the high fields. In contrast to regular core-collapse supernovae, driven by neutrinos, magnetic fields and rotation can influence the behavior of the core collapse in the final evolutionary stage of massive stars. Extensive sets of collapse simulations have been undertaken with varying (parametrized) rotation rates and magnetic fields for the pre-collapse conditions, leading in extreme cases to polar jets and central magnetars.

The magneto-rotational mechanism, proposed in the 1970s \citep{LeBlanc1970}, relies on the extraction of rotational energy from the core via the magnetic field. Therefore, rapid rotation of the iron core is necessary (but not easy  to obtain in simulations up to collapse), as well as an amplification of the magnetic field by rotational winding and/or the magneto-rotational instability \citep[MRI,][]{chandrasekhar60,balbus98}, which was investigated in this environment e.g. by \cite{Obergaulinger2009}. After the core  bounce, the strong magnetic pressure gradient launches jets along the rotational axis \citep{Burrows07,Takiwaki09,Winteler.Kaeppeli.ea:2012,Moesta14,Obergaulinger14}. Initial hopes by \cite{Winteler.Kaeppeli.ea:2012}, were based on a high pre-collapse rotation with a period of 2s at a 1000km radius and a magnetic field in the z-direction of $5\times 10^{12}$ Gauss. Then, only 23 to 30 ms after core bounce the magnetic field is wound up and leads to the ejection of polar jets. The magnetic pressure is a factor 10-100 higher in the jets than the gas pressure. Ejected matter had a (relatively) low entropy of about 10$k_B$ per baryon, not yet heated strongly by outstreaming neutrinos. This is the main idea behind this scenario: the matter, neutronized via electron capture during collapse, has to be ejected fast before neutrino interactions can raise $Y_e$ and lead to a weaker r-process. More realistic simulations with smaller initial magnetic fields, where the magnetic field enhancement takes place during the emerging explosion via the MRI, lead to later ejection and less neutron-rich matter. The outcome is a weak r-process, which extends beyond the second r-process peak at $A=$130 and also produces Eu, but in much smaller than solar r-process proportions.
This is in line with recent 3D similations by \cite{Reichert.Obergaulinger.ea:2023}, \cite{Reichert.Bugli.ea:2024}, \cite{Zha.Mueller.Powell:2024}, and \cite{Prasanna.Coleman.ea:2025}. 
\begin{figure}
    \centering
    \includegraphics[width=0.8\linewidth]{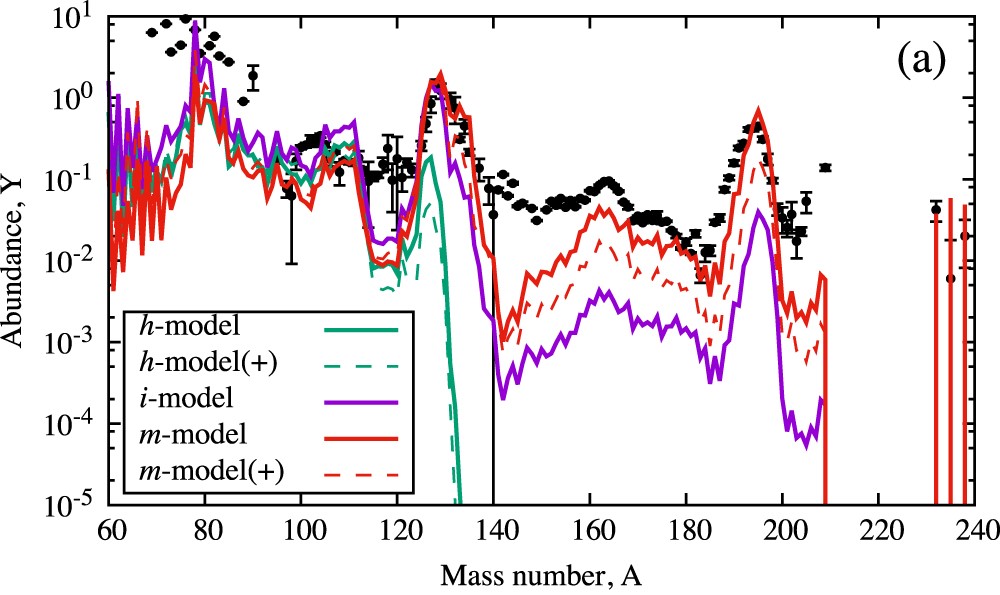}
    \caption{Final abundances, especially with respect to the strength of an r-process, from the parametrized magneto-rotational models by \cite{Nishimura.Sawai.ea:2017} compared to solar r-process abundances \citep{arlandini99}. Notice the variations from the weakest (h) to the strongest (m) magnetic field effects due to the variations chosen for the neutrino luminosity effect in comparison to magnetic fields - the most likely option is probably (i), see also \citep{Reichert.Obergaulinger.ea:2023,Reichert.Bugli.ea:2024}); reprinted with permsission from \cite{Nishimura.Sawai.ea:2017}, copyright by the AAS.}
    \label{fig:Nishimura172}
\end{figure}

Fig.~\ref{fig:Nishimura172} from the study by \cite{Nishimura.Sawai.ea:2017} shows the variation in possible r-process outcomes. The results are based on a 2D axis-symmetric simulation utilizing an initial dipole magnetic field with a central value of $2\times10^{11}$~Gauss, i.e about a factor of 5 weaker than in \cite{Winteler.Kaeppeli.ea:2012}.
The simulation resolves the effect of the MRI, which can enhance the field during collapse and explosion, dependent upon  the rotation. Considering the results from \cite{Reichert.Obergaulinger.ea:2021}, \cite{Reichert.Obergaulinger.ea:2023} and \cite{Reichert.Bugli.ea:2024}, an obvious conclusion is that magneto-rotational supernovae, leading also to highly magnetized neutron stars (magnetars), are only the site of a weak r-process. 

In the case of the {\it{i-model}} this would lead to an Eu production in the range from ${5}\times 10^{-7}$M$_\odot$ to $3\times 10^{-6}$M$_\odot$ with related Fe ejecta from $8\times 10^{-2}$M$_\odot$ to $3\times 10^{-2}$M$_\odot$ \citep[see Fig.5 in][]{Nishimura.Sawai.ea:2017} (average values in the i-model correspond to an Eu/Fe mass ratio in the ejecta of $6\times 10^{-6}$-$10^{-4}$ or Eu/Fe abundance ratios of $2.2\times 10^{-6}$ to $3.7\times 10^{-5}$, with typical mass numbers $A_{Eu}=153$ and $A_{Fe}=56$). These predictions will be compared later with observational numbers in the section on observations. As seen in Fig.\ref{fig:Nishimura172}, magneto-rotational supernovae produce elements beyond the second r-process peak with $A>130$, including Eu, but in largely varying amounts. Therefore, the average production can also be lower than the values given above, but it should also be stressed that this r-process site comes with a co-production of Fe and Eu.

While magnetars form in more than 10\% of core collapse events \citep{Beniamini.Hotokezaka.Horst.ea:2019}, a fraction which could even be as high as 50\% \citep{Pardo.Rea.ea:2026} when extending the magnetic field strength down to $10^{14}$G, this scenario does not necessarily lead to a large-scale arrangement of magnetic fields with jet formation and r-process ejecta in all cases. Thus, not each magnetar is the result of a magneto-rotational supernova explosion and could be formed in neutrino-dominated SNe as well \citep[see the range of explosion strengths shown in Fig.5 of][]{Nishimura.Sawai.ea:2017}. Therefore, the MRSN fraction of all CCSNe will probably not be larger than about 10\% (see discussion in later sections).

Recent investigations on magnetar giant flares \cite{Patel.Metzger.ea:2025a,Patel.Metzger.ea:2025b}, occurring on timescales of 1000 to 10 000 years after the supernova explosion, show that these events also produce a weak r-process with in total about $10^{-6}$M$_\odot$ of r-process ejecta. Such giant flares occur during the cooling phase of the magnetar and can repeat several times. This material is merged with the preceding supernova ejecta in the same supernova remnant.
\subsubsection{Collapsars/Hypernovae}
\label{sec:colhyp}
There exist explosions with a supernova appearance, despite the formation of a central black hole. The mechanism which enables this is due to (similar as in the previous subsubsection) fast rotation and, eventually,  strong magnetic fields. It involves polar jet ejection, accompanied by long-duration gamma-ray bursts (GRBs) and accretion disks with possible outflows \citep[for reviews on GRBs, and especially to the distinction between long and short bursts, related to different origins see, e.g.][]{Piran:2004,Piran.Bromberg.ea:2013,mckinney13,mckinney14,ruiz16,murguia17,ruiz19,Gottlieb.Lalakos.Bromberg.ea:2022,Gottlieb.Metzger.ea:2025};  for the physics of accretion disks see the classic textbook by \cite{Frank.King.Raine:2002}.

The collapsar origin for such GRB hypernovae has been
suggested in a number of investigations by the Woosley group \citep{Woosley:1993,macfadyen99,macfadyen01,Woosley.Bloom:2006}, as well as the Nomoto group, \citep[summarized in][]{Nomoto:2017}. The question is how the average kinetic energy found in observations to be $\approx 2.5\times 10^{52}$erg, an ejecta mass of $\approx$ 6 M$_\odot$ with a nickel mass of $\approx$ 0.4 M$_\odot$ for these GRB hypernovae \citep{Cano.ea:2017}, can be attained in numerical simulations. \cite{Heger.Mueller.Mandel:2023} discuss
how the formation of the relativistic GRB jets involves the extraction of rotational energy from the black hole via electromagnetic fields \citep{Blandford.Znajek:1977} or via hydromagnetic flows from rotating accretion disks \citep{Blandford.Payne:1982}.

\begin{figure}
    \centering
    \includegraphics[width=0.45\linewidth]{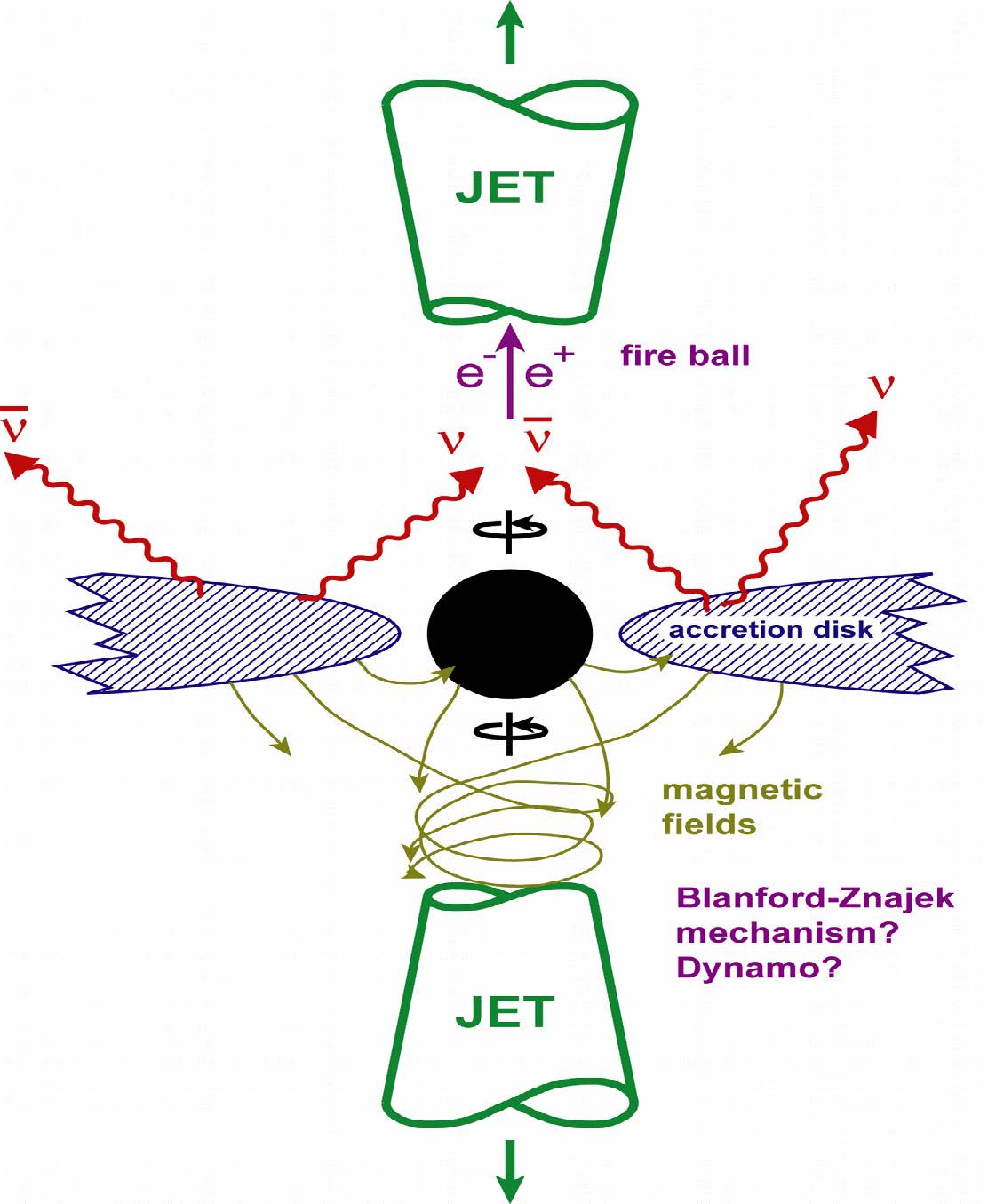}
    \includegraphics[width=0.45\linewidth]{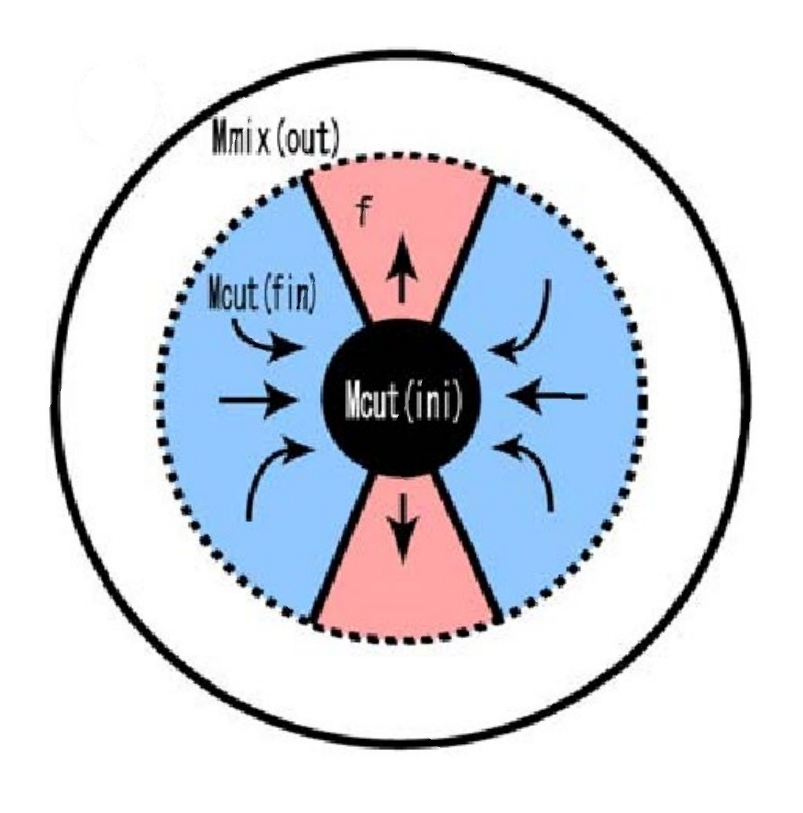}
    \caption{Jet ejection due to extraction of rotation energy from the central black hole and infalling matter from a rotating accretion disk.
    Left: listing the involved physical mechanisms (image courtesy of A. Heger), right: a variation of the mixing-fall back model. Within the dotted cube material is mixed and a fraction $f$ ejected via jets only within a given opening angle as suggested by \cite{Maeda.Nomoto:2003}.
    Image adapted from \cite{Nomoto.Tominaga.ea:2006}.}
    \label{fig:jetSNe}
\end{figure}

Fig.~\ref{fig:jetSNe} shows two views of these effects from the Woosley and Nomoto schools, indicating how massive stars, which are rotating and forming black holes, can experience a final explosions. The infalling matter
forms an accretion disk. This accretion disk
releases gravitational energy (a large fraction of the disk rest-mass for rotating Kerr black holes).
The released energy allows for polar jet ejection, which, combined with winds off the hot disk, leads to the explosive ejection of stellar material.

The question what kind of nucleosynthesis occurs in such black-hole-forming hypernovae/collapsars is complex and related to 
the ejecta composition emerging in the jets, as well as possible accretion disk outflows. It requires close to self-consistent simulations, including magneto-hydrodynamics, gravity and neutrino transport, and the equation of state of highly compressed matter, which controls the transition to black holes in relativistic calculations. Relativistic magneto-hydrodynamic simulations (also including weak interactions and neutrino transport) have been undertaken by \cite{Gottlieb.Lalakos.Bromberg.ea:2022} within a 3D GRMHD treatment. They find a bipolar jet ejected from the accreting (rotating) Kerr black hole, breaking out from the collapsing star (see also \cite{Janiuk.Sapountzis:2018,Fujibayashi.Lam.ea:2024,Shibata.Fujibayashi.ea:2024}) with an overall energy of the order $10^{52}$erg. This is in line with the conditions assumed by \cite{Nomoto:2017,Leung.Nomoto.Suzuki:2023}, indicating that large amounts of $^{56}$Ni are produced. 

Another question is how heavy elements, and possibly r-process matter, can originate from collapsars. 
The ejected matter can have passed through different types of conditions with the following possibilities (a), (b), and (c): (a) matter from stellar burning shells with $Y_e$ close to 0.5, not yet processed during the collapse and explosion, (b) matter which passed through the neutrino cooling phase, i.e., being neutronized - down to $Y_e<0.25$ - mainly by electron captures (with high Fermi energies) in high density environments, and (c) matter being heated by neutrino irradiation, where capture of electron neutrinos and antineutrinos of similar energies on neutrons and protons favors the increase of $Y_e$ up to 0.5 or even beyond, due to the $\approx$1MeV mass difference between neutrons and protons. However, antineutrino captures on protons which sufficiently high energies can also lead to neutrons. The previous paragraph focused on shocked matter of type (a), which leads to $^{56}$Ni production.
Jets can also be launched,  including matter falling in  towards the black hole having undergone conditions (b) at high densities in the inner region of accretion disks. A number of works utilized this scheme \citep{Fujimoto.Nishimura.Hashimoto:2008,Ono.Hashimoto.ea:2012,Nakamura.ea:2013}, proposing that low $Y_e$ conditions are attained in outflows from relativistic collapsar jets. On the other hand, predictions for accretion disk outflows exist with apparently quite different outcomes. 
Pioneering nucleosynthesis studies \citep{Surman.McLaughlin:2004,McLaughlin.Surman:2005,Surman.etal:2006} pointed out that neutrinos can play a critical role, reducing - via conditions of type (c) in neutrino-driven winds - the neutron-richness of outflowing matter and therefore reducing the possibilities for an r-process (see Eq.(\ref{eq:yeeq}) discussed with respect to ``regular''  neutrino-driven supernova explosions). Alternatively, as a follow-up for such initially proton-rich conditions, a scenario has recently been proposed, which - under specific conditions in high-entropy winds and for enhanced neutrino luminosities and fast dynamical timescales - can convert excess protons to neutrons by antineutrino capture \citep[as discussed above and similar to the $\nu$p-process][]{Froehlich.Martinez-Pinedo.ea:2006}, introducing a neutron-capture reaction flow which can even produce lanthanides \citep[the so-called $\nu$i-process][]{Wang.ea:2026}.

On the other hand, \cite{Siegel.Barnes.Metzger:2019} performed 3D GRMHD simulations that create neutron-rich winds, which arise from MRI-driven disks, occurring even for weak initial magnetic fields of the progenitor star. These disk winds are driven by thermal pressure gradients \citep[see also][]{Janiuk:2014,Plonka.Janiuk:2026}.
Based on these mechanisms the disk midplane settles inside of an inner radius of $\sim 10\, GM_{\rm BH}/c^2$ with electron fractions of $Y_e\approx 0.1$ \citep{Beloborodov:2003}. This occurs once the accretion rates exceed an ``ignition value $\dot M_{ign}$'' which depends on the BH spin \citep{chen07}, also determining the corresponding accretion rates needed to power long GRBs \citep{Lee.Ramirez-Ruiz:2007}. This self-regularization to low $Y_e$-values in the disk midplane is found in full-fledged numerical (magneto-) hydrodynamic simulations, see e.g. \cite{Siegel.Metzger:2018} and \cite{Fernandez.Tchekhovskoy.ea:2019}. Generally a full r-process can be obtained for $Y_e<0.25$.

\begin{figure}
    \centering
    \includegraphics[width=0.7\linewidth]{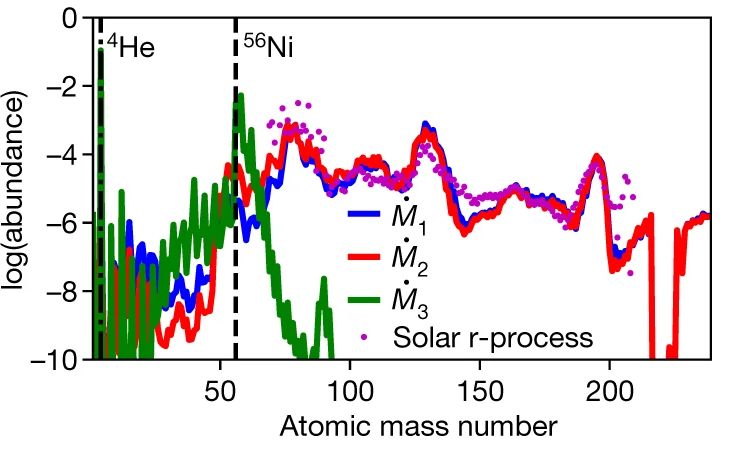}
    \caption{Abundance distributions of nuclei synthesized in the
disk outflows at three different accretion stages (dots represent
the observed Solar System abundances) from \cite{Siegel.Metzger:2018} and \cite{Siegel.Barnes.Metzger:2019}. Above $\dot M_{ign}$, a complete r-process up to
atomic mass numbers of around 195 is obtained ($\dot M_1$, $\dot M_2$), whereas below $\dot M_{ign}$
a rapid transition to outflows rich in $^{56}$Ni and $^4$He is observed ($\dot M_3$); image from \cite{Siegel.Barnes.Metzger:2019}.}
    \label{fig:Siegelrprocess}
\end{figure}

In principle, several phases of the disk formation need to be considered: an initial neutrino-cooling dominated accretion flow (NDAF) phase leads to small $Y_e$'s and is consistent with  condition (b) discussed above, when viscous heating is balanced by neutrino cooling. Neutrino cooling depends on temperature and density (or electron degeneracy). 
The GRMHD investigations by \cite{Siegel.Metzger:2017,Siegel.Metzger:2018,Miller.Sprouse.ea:2020,Fernandez.Tchekhovskoy.ea:2019} agree that a large fraction ($\sim 40\%$) of the initial torus mass becomes unbound, but at this time there is no full agreement about the resulting $Y_e$ and composition of the ejecta. For example, \cite{Fernandez.Tchekhovskoy.ea:2019} find $Y_e$ values around 0.12, those of \cite{Siegel.Metzger:2018} peak around $\sim 0.14$ (see Fig.\ref{fig:Siegelrprocess}), while \cite{Miller.Sprouse.ea:2020} and \cite{Shibata.Fujibayashi.Wanajo.ea:2025} find a broad distribution between 0.2 and 0.4/0.5.

The neutrino-cooling-dominated accretion flow (NDAF) phase can be followed by another phase of the disk accretion, where neutrino cooling no longer dominates and
viscous heating becomes dominant. This can lead to an advection dominated accretion flow (ADAF), driven by turbulent angular momentum transport within the disk, possibly accreting neutron-rich matter into the black hole. This leads to a heated thick disk, expansion to lower densities, and the production of  outflows. The main question is whether, and how or when,  this transition takes place.  That in turn  determines whether neutron-rich matter (consistent with a strong r-process) can be ejected early, or matter with $Y_e$ closer to 0.5 is ejected in those outflows. Simulations with a full GRMHD (or even a neutrino-transport $\nu$GRMHD) treatment
eject more neutron-rich matter during  the early NDAF phase \citep{Issa.Gottlieb.ea:2025}, while studies, utilizing an $\alpha$-disk treatment \citep{Just.Goriely.Janka.ea:2022},
find that the ADAF phase dominates and leads only to high $Y_e$ in the ejecta. In all cases, during the late phase of disk accretion  $^{56}$Ni (with $Y_e$$\approx$0.5) is also ejected in non-negligible amounts.
A further issue is whether the neutron-rich inflow from the inner disk, occurring during the ADAF phase,  is fully absorbed in the black hole or partially ejected along the powerful jet resulting from the infall \citep{Gottlieb.Lalakos.Bromberg.ea:2022}? While there is not much matter in the jet itself, it can accelerate matter from the innermost disk material, as previously  suggested by \citep{Fujimoto.Nishimura.Hashimoto:2008,Ono.Hashimoto.ea:2012,Nakamura.ea:2013}. Contrary to MR supernovae, where the jets can be weak and undergo kink instabilities, collapsars develop powerful jets.
In this respect it might also be of interest that \cite{Mumpower.Lee.ea:2025} could show that typical stellar baryonic material can become inundated with neutrons in situ via hadronic photoproduction, which could take place in collapsars, containing substantial flux of high-energy photons and would be favorable for neutron-capture nucleosynthesis.

In order to come up with typical predictions of r-process ejecta masses in collapsars/hypernovae we find the values by \cite{Siegel.Barnes.Metzger:2019}, who predict $>10^{-1}$M$_\odot$ of r-process matter and typically 0.5M$_\odot$ of $^{56}$Ni (decaying to Fe), while \cite{Brauer.ea:2021} predict slightly smaller masses for ejected r-process matter: $7 \times 10^{-2}$M$_\odot$. This translates to about $7 \times 10^{-5}-10^{-4}$M$_\odot$ of Eu, an Eu/Fe mass ratio of $(1.4-2)\times 10^{-4}$ and an abundance ratio of $(5.2-7.4)\times 10^{-5}$, but such numbers need to be confirmed by further studies which also may lead to smaller values, \citep[e.g.][]{Shibata.Fujibayashi.Wanajo.ea:2025}. Given the fact that the solar Eu/Fe ratio is $1.15\times 10^{-7}$ \citep{Lodders.Bergemann.Palme:2025}, it should be pointed out that the predicted Eu/Fe ratios in the ejecta are more than a factor of 100 higher than solar values and therefore (in comparison to solar ratios) the amount of Fe produced in this site is negligible.
Finally, it should be mentioned that \cite{Brauer.ea:2021} predict collapsar events to CCSNe with a ratio of about 1 out of 1000.

\subsection{Compact binary mergers}
\label{sec:nsm}

\begin{figure}[h!]
    \centering
    \includegraphics[width=0.8\linewidth]{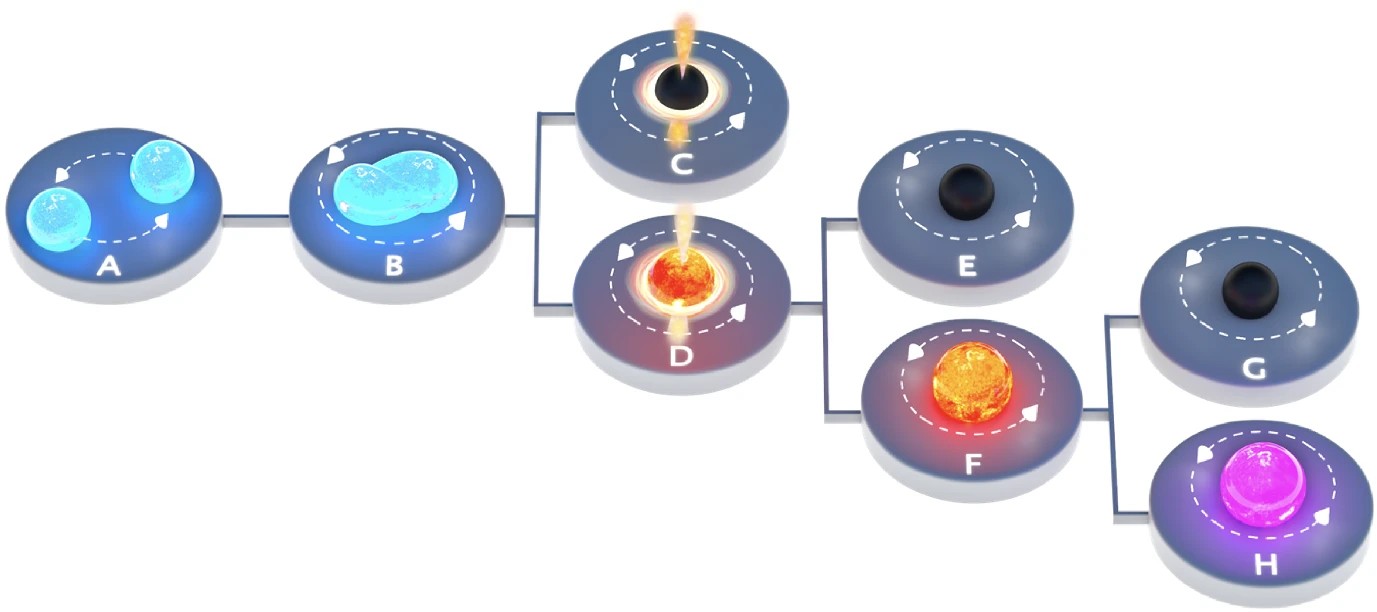}
    \caption{Fate of binary neutrons star mergers, leading (for central objects $>1.5M_{max}$) to prompt black hole formation or otherwise to a rapid differentially rotating neutron star. The latter can be hypermassive ($>1.2M_{max}$), collapsing to a BH within about 1s, supramassive ($>1M_{max}$), collapsing to a BH within about $10^5$s, or be infinitely stable ($<1M_{max}$), with $M_{max}$ being the maximum stable neutron star mass. This evolution also includes early dynamical ejecta, jet formation and accretion disks; image from \cite{Sarin.Lasky:2021}.}
    \label{fig:Sarin}
\end{figure}

Producing the heaviest (r-process) nuclei in neutron star mergers was first suggested by \cite{Lattimer.Schramm:1974} with explorations of expected nucleosynthesis patterns by \cite{Meyer.Schramm:1988} and the suggestion of related gamma-ray bursts by \cite{Eichler.Livio.ea:1989}. The first hydrodynamic studies of such events were undertaken by \cite{davies94} and \cite{Rosswog.Davies.ea:2000} with later nucleosynthesis predictions \citep{Freiburghaus.Rosswog.Thielemann:1999}. A large number of investigations were undertaken before the observation of GW170817 \citep[for a review see e.g.][]{Thielemann.Eichler.ea:2017}. Further reviews and recent results related to the merger events and nucleosynthesis ejecta can be found in \citep{Bauswein.Just.ea:2017,Shibata.Fujibayashi.ea:2017,Radice.Perego.ea:2018a, Horowitz.ea:2019,Shibata.Hotokezaka:2019,Metzger:2019,Burns:2020,Radice.Bernuzzi.Perego:2020,Miller.Sprouse.ea:2020,Sarin.Lasky:2021,Cowan.Sneden.ea:2021,Barnes.ea:2021,Shibata.ea:2021,Nedora.ea:2021, Perego.Thielemann.Cescutti:2022,Just.Aloy.ea:2022,Just.Goriely.Janka.ea:2022,Kullmann.Goriely.ea:2023,Fujibayashi.Kiuchi.ea:2023,Chen.Li.ea:2024,Shibata.Fujibayashi.Wanajo.ea:2025,Janiuk.ea:2026}, related to the possible evolutionary paths shown in Fig.\ref{fig:Sarin}.

Three components of neutron star merger ejecta contribute to the overall nucleosynthesis: (i) dynamical ejecta including compressed and shock heated material from the initial collision,  as well as possibly -- cold -- tidal spiral arm-type ejecta (happening within about 1ms with $Y_e$'s of the order 0.05-0.25 in equatorial ejecta \citep[see e.g.][]{Freiburghaus.Rosswog.Thielemann:1999,Goriely.Bauswein.Janka:2011,Rosswog.Korobkin.ea:2014,Eichler.Arcones.ea:2015} and $Y_e$$\approx$0.25-0.5 in polar ejecta \citep[e.g.][]{Martin.Perego.ea:2018,Radice.Perego.ea:2018b,Kullmann.Goriely.ea:2022}, (ii) winds driven by neutrinos, emitted from the central hot very massive neutron star and the accretion disk \citep[e.g.][]{Martin.Perego.ea:2015,Wu.Tamborra.ea:2017, Just.Goriely.Janka.ea:2022} and potentially also by magnetic fields (within about 100ms with $Y_e$'s also of the order 0.25-0.5 if a stable neutron star remnant is eventually left) and (iii) finally,  mass outflow from the accretion disk \citep[e.g.][]{Wanajo.Sekiguchi.ea:2014,Just.Bauswein.ea:2015,Wu.Fernandez.Martinez.ea:2016,Just.Aloy.ea:2022}, for hypermassive neutron stars within about 1s and $Y_e$'s in the ejecta of the order 0.1-0.4, combined with a short-duration gamma-ray burst - GRB - after black hole formation. A common feature of these scenarios is that matter reaches a nuclear statistical equilibrium (NSE) distribution with $Y_e$ given by weak reactions or in the cold dynamical ejecta component by beta equilibrium in the cold neutron stars before merger. Low $Y_e$'s in the central parts of the accretion disk torus require high densities (and high electron Fermi energies) as they are due to electron capture on protons. The conditions discussed here are given in detail in Fig.31 of \cite{Cowan.Sneden.ea:2021}, see also \cite{Rosswog.Feindt.ea:2017}. 
\begin{figure}[h!]
    \centering
    \includegraphics[width=0.65\linewidth]{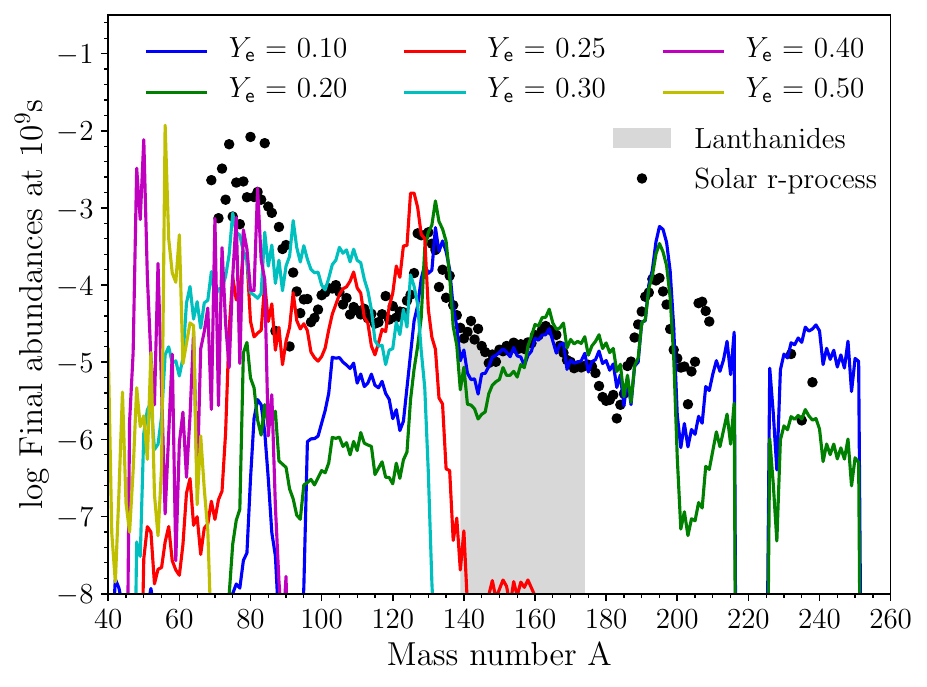}
    \caption{Abundances as a function of mass number $A$ (left panel) at 10$^9$s after the merger of two neutron stars of 1.35M$_\odot$ for trajectories characterized by $s \approx 11k_{\rm B}{\rm baryon^{-1}}$ and $\tau \approx 11~{\rm ms}$, but for different initial $Y_e$'s, computed using the SkyNet nuclear network \citep{Lippuner.Roberts:2017}; image reproduced with permission from \cite{Perego.Thielemann.Cescutti:2022}, copyright by Springer Nature. Black dots represent the Solar $r$-process residual, as reported by \cite{Prantzos.Abia.ea:2020}.}
    \label{fig:rabund-Aye}
\end{figure}

\begin{figure}[h!]
    \centering
    \includegraphics[width=0.68\linewidth]{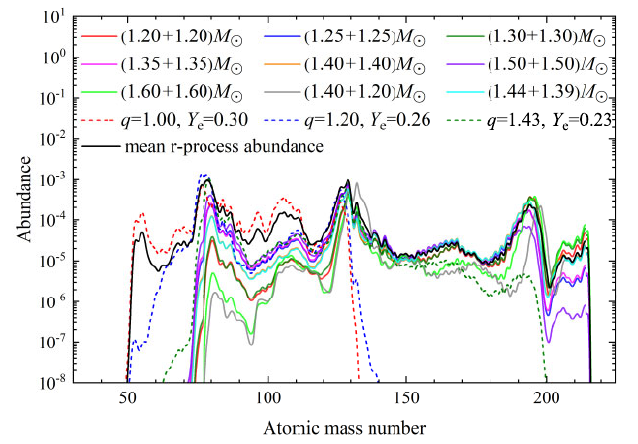}
    \caption{The resulting r-process abundances for dynamical ejecta (solid lines) and disc wind ejecta (dashed lines) in BNS mergers. Astrophysical inputs for r-process nucleosynthesis simulations for dynamical ejecta correspond to the binary masses listed on the side of the color coding, the disk wind ejecta are indicated by the mass ratio of the binary system $q$ and the electron fraction $Y_e$. The black line represents a weighted superposition of all shown contributions. Figure from \cite{Chen.Li.ea:2024}; copyright by authors.}
    \label{fig:rabund-Chen}
\end{figure}

Thus, the central merged object (even if it is beyond the stable neutron-star mass limit) can initially be supported by thermal pressure and rotation, formimg a hypermassive neutron star that blows off a neutrino-powered wind, as in core-collapse supernovae, preferentially in the axis direction. After the formation of the black hole, an accretion disk forms that leads to axial jets (causing a gamma-ray burst with very large Lorentz factors $\Gamma$) and accretion disk outflows. The tidal (almost pristine) very neutron-rich ejecta and the accretion disk outflows form the heaviest elements. The neutrino-powered wind increases the (initially very low) $Y_e$ from values $<$0.1 up to possibly $>$0.3-0.4 via neutrino captures on neutrons ($\nu+n \rightarrow p+e^-$). This would cause only a weak r-process and less massive nuclei.

Fig.\ref{fig:rabund-Chen} summarizes a comprehensive set of abundance predictions for a variety of binary neutron star systems, with symmetric systems varying from 1.2 to 1.6M$_\odot$ and also two asymmetric systems. The term "dynamical ejecta" refers to  matter ejected in the equatorial plane with typical initial $Y_e$'s, entropies, and expansion velocities, taken from \cite{Radice.Perego.ea:2018b} and \cite{Nedora.ea:2021}, while "wind ejecta" refers to ejected matter off the equatorial plane. The resulting abundance patterns are labeled with their binary masses or mass ratios \citep[for detailed conditions see Table 1 in][]{Chen.Li.ea:2024}. 

The higher density of electronic states in the heaviest elements (lanthanides and actinides)  causes a higher impact of photon scattering to lower energies, and photonic radiation transport results in a red appearance of the hot object, while the intermediate to light heavy elements would cause the object to appear blue. 
Combining the types of ejecta discussed above leads to an overall agreement with the solar system r-process pattern, but due to the different ejecta directions (equatorial and perpendicular) this could lead to differing observational appearances, if both ejecta types could be observed independently.
This will also be an important topic in the following section related to observations. 

At the end of this section we want to also provide a typical value for r-process ejecta from neutron star mergers with a total ejected r-process mass in the range ($1-4)\times 10^{-2}$M$_\odot$ \citep{Cote.Belczynski.ea:2017,Cote.Fryer.ea:2018,Fujibayashi.Kiuchi.ea:2023}. Assuming  a solar r-process abundance distribution, this corresponds to $10^{-5}$-$4\times 10^{-5}$M$_\odot$ in Eu. However, a different range of $(3-15)\times 10^{-6}$M$_\odot$ is given by \cite{Cote.Fryer.ea:2018}, i.e., predicting somewhat smaller values. Neutron star mergers can produce Fe, but in negligible amounts, leading to Eu/Fe ratios on  the order $10^5\times$ solar (Zewei Xiong, private communication). Thus, if we discuss Eu/Fe ratios in the observational sections on low-metallicity stars, it should be clear that in patterns imprinted by neutron star mergers, most of the Fe has to come from preceding core-collapse supernova contributions.
The total neutron star merger rate in comparison to the CCSN rate is estimated to be of the order $4.5\times 10^{-3}$ \citep{Brauer.ea:2021}. An error range of 
$(2.3-5.1)\times 10^{-3}$ is given in \cite{Grunthal.Kramer.ea:2021} based on binary neutron star observations. If binary neutron star mergers can be identified with short gamma-ray bursts, their event rate should be identical with those of sGRBs, estimated (in volumetric rate densities) to be of the order $160^{+200}_{-100}$Gpc$^{-3}$yr$^{-1}$ \citep{Dichiara.ea:2020} or even $156-3200$ Gpc$^{-3}$yr$^{-1}$ \citep{Liu.Zhang.Zhu:2021} and $303^{+1580}_{-300}$Gpc$^{-3}$yr$^{-1}$ \citep{Howell.ea:2025}. The higher portion of these rates shows a clear overlap with the numbers from \cite{Grunthal.Kramer.ea:2021}, which can be translated into a universal rate of $300-650$ Gpc$^{-3}$yr$^{-1}$. On the other hand, the statistics of gravitational wave detections of (up to now) only two observed neutron star mergers in the Gravitational Wave Transient Catalogue GWTC-4 \citep{Abac.ea:2025} translates to $28-300$Gpc$^{-3}$yr$^{-1}$ neutron star merger events, a relatively low number of events, overlapping, however, with the previously presented rates \citep[for a further discussion see][]{Fishbach.ea:2026}. In the later discussion on nucleosynthesis in the early Galaxy we will proceed with assuming rates of $>3\times 10^{-3}$ neutron star mergers per CCSN.

\subsection{Other suggested contributions}
\label{sec:other}
In addition to binary neutron star mergers, simulations of mergers of neutron stars with black holes have been performed in detailed simulations \citep{Desai.Metzger.Foucart:2019,Curtis.Miller.ea:2023,Wanajo.Fujibayashi.ea:2024,Kawaguchi.ea:2024}. These authors predict approximately 0.07M$_\odot$ of r-process ejecta, which would result in approximately  $7\times 10^{-5}$M$_\odot$ in Eu. \cite{Wanajo.Fujibayashi.ea:2024} predicted that black hole-neutron star mergers lead to a boost in actinide abundances with $Y_e$ values going down to 0.05 (see their Fig.1), while \cite{Farouqi.ea:2022} suggested the same for collapsars. These specific results depend on the ejection of low $Y_e$-matter (see Figs.\ref{fig:meng-ru} and \ref{fig:rabund-Aye}). 
\cite{Harry.Hoy:2026} estimate the occurrence frequency of neutron star - black hole mergers to be smaller than 1 event per 1000 CCSNe. The recent GWTC-4 catalogue \citep{Abac.ea:2025} gives neutron star - black hole merger rates in the range of $9.1-84$Gpc$^{-3}$yr$^{-1}$, which is smaller than the above mentioned value, but results in a similar ratio of black hole - neutron star mergers to binary neutron star mergers.

Further suggestions include mergers of neutron stars with the cores of massive stars during a commmon envelope phase \citep[so-called common envelope jet supernovae CEJSNe,][]{Grichener.Kobayashi.Soker:2022,Jin.Soker:2024,Grichener:2025}, predicting masses of Eu ejecta in the range $1-3\times 10^{-5}$M$_\odot$, when employing analytical approaches. These numbers are comparable to those found in the previous discussion of neutron star mergers, but they need to be confirmed with full GRMHD simulations. The occurrence frequency of these events will be discussed in later sections. 

Mergers of neutron stars with white dwarfs have been discussed with respect to their nucleosynthetic outcomes \citep{Metzger:2012,Margalit.Metzger:2016,Fernandez.MetzgerWD:2013,Fernandez.Metzger:2019} and were also suggested as sites for r-process ejecta \citep{Chrimes.ea:2025}.
This results in accretion-induced collapse (AIC) of white dwarfs in binary systems or merger-induced collapse (MIC), leading to neutron star and/or magnetar formation, as well as the ejection of matter. This was  recently discussed by \cite{Batziou.Janka.ea:2025} as possible sources of an r-process. The average $Y_e$ of the ejecta is in the range 0.42 to 0.5 (see their Fig.8), and - dependent upon rotation - smaller amounts of matter (of the order of $2\times 10^{-3}$M$_\odot$) with a $Y_e$ range 0.26-0.4 can be part of the ejecta.  According to Fig.\ref{fig:rabund-Aye} this would produce matter up to the second ($A\approx 130$) r-process peak, i.e., this scenario is a candidate for a weak r-process. 
Assuming very high initial magnetic fields in excess of $10^{11}-10^{12}$G \citep{Cheong.Pitik.ea:2025}, i.e., 3-4 orders of magnitude higher than found in any white dwarf observations and similar to the very optimistic initial assumptions for magneto-rotational supernovae by \cite{Winteler.Kaeppeli.ea:2012},
even $Y_e$ values down to 0.1 could be attained (see their Fig.5). A further study by \cite{Pitik.Radice.ea:2026} underlines this effect. They argue that such unrealistically high initial fields have to be chosen to mimic MRI effects during the collapse, which would then amplify initially smaller fields to such strengths. This effect is also known in magneto-rotational supernovae. However, the necessary field amplification by the MRI delays the explosion, giving way to neutrino radiation effects which increase $Y_e$ again and would reduce the strength of the r-processing.
We will discuss these varieties of white dwarf accretion or mergers and the possible detectable outcomes in the next section on observations. 

Further suggestions include subminimal-mass neutron stars. Based upon ideas by \cite{Clark.Eardley:1977}, stable mass transfer in neutron star binaries of unequal mass can reduce the mass of the lower-mass companion, proceeding towards the minimum neutron star mass. Such scenarios predict a tidal disruption and explosion of the neutron star at or below the minimum stable mass \citep[see e.g.][and references therein]{Blinnikov.ea:2022}, however, Fig.5 of \cite{Rosswog.Piran.Nakar:2013} should also be considered for the survival chances of the neutron star. Such an event is expected when a large ratio for the masses of the two compact objects exists, which would be the case for neutron star - black hole pairs \citep{Martineau.ea:2026}. \cite{Yip.Leung.ea:2023} and \cite{Ignatovsky.ea:2023} predict r-process nucleosynthesis ejecta from such explosions, which would add an additional formation channel beyond the previously discussed ones. The authors state that the ejected mass of r-process elements is as high or higher than those of neutron star mergers. Therefore, their contribution depends strongly on the event rates and how they compare with those of neutron star mergers. In principle, because neutron star - black hole mergers always show a large mass ratio of the binary companions, such events could be highly correlated with neutron star - black hole mergers.

\section{r-Process Observations}
\subsection{r-process elements seen in astrophysical explosions}

While r-process abundances in the solar system have been determined with increasing precision for some time (see Sections \ref{sec1} and \ref{sec2}), low abundances of heavy elements in astrophysical explosions, combined with Doppler broadening of lines, have made it difficult to observe these elements directly in such events.
\subsubsection{Compact Binary Mergers}
It was not until 2017 that direct evidence was found for a production site of r-process elements. In a fortuitous observation in 2017 gravitational waves, followed by light emission (i.e., a kilonova event), indicated that a neutron star merger had taken place.
The follow-up of the gravitational wave event GW170817 \citep{Abbott.Abbott.ea:2017} revealed strong electromagnetic emission in the aftermath of the merger \citep{kasliwal17,evans17,Villar.Others:2018,drout17,Wu.Barnes.ea:2019,Miller.Ryan:2019,Kasliwal.Kasen.ea:2022} and showed, in particular, the expected signatures of an r-process-powered kilonova. The decay of its bolometric lightcurve agreed well with the expectations for radioactive heating rates from a broad range of r-process elements \citep{Metzger.Martinez.ea:2010,rosswog18,Zhu.Wollaeger.ea:2018,Metzger:2019,Wu.Barnes.ea:2019,Barnes.ea:2021,Lund.ea:2023,Just.ea:2023,Kawaguchi.ea:2023,Kawaguchi.ea:2024}.
The kilonova duration, brightness, and color provided critical information about the composition, amount, and velocity of the matter ejected. Observations of the 2017 kilonova (AT2017gfo) showed that it was initially blue, indicating the synthesis of elements lighter than barium (Z = 56) \citep{Kasen.Metzger.ea:2017}. These elements have a low density of atomic levels and make the medium less opaque, allowing light to escape (decouple from matter) earlier and without having lost much energy. The presence of lighter r-process elements was also confirmed by the direct observation of strontium (Z = 38) in the spectra of AT2017gfo \citep{Watson.Hansen.ea:2019}. After a few days, the light of AT2017gfo turned from blue to red, pointing to the presence of lanthanides and actinides. 
In summary, there is strong evidence that this neutron star merger event has produced at least a broad, and maybe the whole, r-process abundance range. However, based on the observed lanthanide fraction $X_\mathrm{La}$, \cite{Waxman.Ofek.ea:2018}, \cite{Ji.Drout.Hansen:2019}, and \cite{Gillander.Flors.Ferreira:2025} found that at least for the neutron star merger GW170817, a typical solar r-process pattern is not found. This is consistent with recent predictions for additional cases \cite{Holmbeck.Andrews:2024}, but see also \cite{Vieira.Ruan.ea:2026}.

Further kilonova observations related to GRB 211211A \citep{Yang.ea:2022} and GRB 230307A \citep{Levan.ea:2024} seem to support light-curves powered by the decay of heavy neutron-rich ejecta, and - in fact - in the spectrum for  GRB 230307A the second r-process peak element Te has been observed as well. However, both of these events belong to the class of long-duration GRBs, usually related to the collapse of very massive stars/hypernovae. It has to be verified whether binary neutron star mergers, possibly with stronger magnetic field configurations, or other objects like neutron star - white dwarf / neutron star - massive star, or neutron star - black-hole mergers are the origin of such events \citep[see e.g.][]{Chrimes.ea:2025}.

\subsubsection{Magnetorotational supernovae and giant magnetar flares}

There has been a recent observation of a magnetar giant flare with a previously unexplained hard gamma-ray signal seen in the aftermath of the famous 2004 December giant flare from the magnetar SGR 1806-20.
\cite{Patel.Metzger.ea:2025a,Patel.Metzger.ea:2025b} could show that the MeV emission component, rising to a peak at around 10 minutes after the initial spike, before decaying away over the next few hours, is direct observational evidence for the synthesis of $\approx10^{‑6}$ M$_\odot$ of r-process elements. This compares with about the same amount of ejected mass in Eu alone that is  predicted for magneto-rotational supernovae (see Section \ref{sec:MHD}). However, since several (up to 10 of such) flares can take place during the cooling phase of the magnetar, the total contribution could be as high as $\approx10^{‑5}$ M$_\odot$ of r-process elements. This provides evidence that some small fraction of the Galactic r-process abundances might be produced in such an event. These events occur on timescales of 1000 to 10,000 years after the preceding magnetar formation, following the magneto-rotational supernova explosion, and produce a weak r-process as well. The synthesized  nuclei are merged with the supernova ejecta in the same supernova remnant and thus are consistent with our discussion of magneto-rotational supernova contributions in Section \ref{sec:MHD}.

\subsection{r-process contributions in the early Galaxy}
\label{sec:earlygal}
Ancient, metal-poor (or even metal-deficient so-called PoP I) stars were formed from Galactic gas in the early universe that consisted entirely of big bang H, He, and Li. When Population II stars were formed, they could already contain heavier elements, ejected into the interstellar medium by the first exploding objects, and being imprinted with the element composition of their ejecta. Such low-metallicity, very metal-poor (VMP) and extremely metal-poor (EMP), stars are crucial for studying early Galactic evolution. Therefore, a number of observing programs have been set up in order to understand element abundance patterns inherited from the first and earliest stellar generations. These include the HK and Hamburg/ESO surveys \citep{Beers.Christlieb:2005} and the Hamburg/ESO R-process Enhanced Star (HERES) survey \citep{Christlieb.ea:2004}, aimed at studying low-metallicity stars with [Fe/H]$<$-2.5. The 4MOST survey \citep{Storm.ea4MOST:2025} covers a large sample of low-metallicity stars, but only for a limited number of elements up to Eu. The transition from low-metallicity stars to medium-metallicity stars in the range -2.5$<$[Fe/H]$<$-1.5 has been analyzed in the "Measuring at Intermediate Metallicity Neutron-Capture Elements" (MINCE) project \citep{CescuttiMINCEI:2022,FrancoisMINCEII:2024}. 
A very ambitious search for r-process elemental abundances by the "R-Process Alliance" \citep[an interdisciplinary collaboration of astronomers and nuclear physicists][]{Hansen.Alliace:2018}, has now produced  five data releases \citep{Bandyopadhyay.Alliance:2024},  and is aiming soon for the release of the abundances of 2000 metal-poor stars. 
A further program, focusing on the chemical evolution of r-process elements in stars up to metallicities of [Fe/H]=-1.5, is known as CERES \citep{APULS.CERES:2025}, whose goals are a careful analysis of the universality of the r-process, i.e., looking for stellar abundance patterns of the third r-process peak (containing e.g., Os, Ir, Pt).

Additional discussion related to applying all these data to the understanding of  the Galactic chemical evolution of r-process elements, and possibly identifying the sites of their origin, will be given in the next sections. Here we give a short broad review before addressing more details.

It has been known for some time that early in the history of the Galaxy, element production - even for those elements typically thought of as s-process elements like Ba - occurs predominantly as a result of the r-process 
\citep[see e..g][]{sneden83,Sneden.Cowan.Gallino:2008}. In terms of metallicity [Fe/H] the s-process starts to set in in the range -2.5 to -2. At earlier times (i.e., lower metallicities) in Galactic evolution the r-process dominates the production of heavy elements. 
This points to the fact that ejecta from relatively fast evolving astrophysical objects, and their explosions, dominate the early Galaxy. At that time the interstellar medium is still almost pristine, i.e. consisting of big bang abundance patterns, and only the first stellar explosions have contributed to matter forming the next generation of stars, which show in their surface abundances the composition with which they were born. Among the objects discussed in the previous section, regular core-collapse supernovae, magneto-rotational supernovae (plus their magnetar giant flares) and collapsars/hypernovae, i.e. all events related to the core collapse of massive stars, clearly fulfill this requirement for a rapid evolution before explosive ejection of matter. It is not obvious whether compact binary mergers belong to this category as well; this will be discussed in the next section.

\begin{figure}[h!]
\centering
\includegraphics[width=0.76\textwidth]{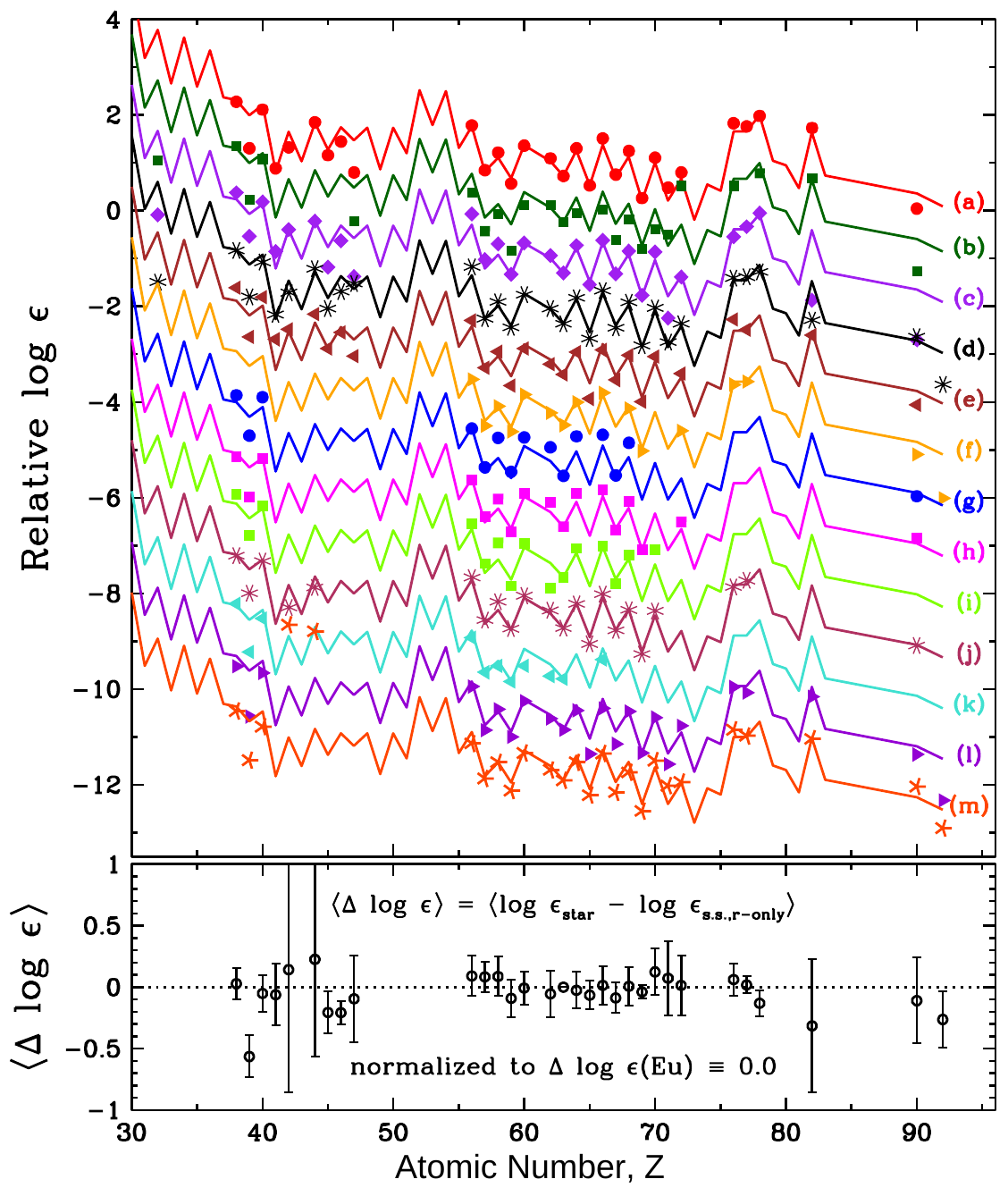}
\caption{Top panel: neutron-capture abundances in 13 r-II stars
(points) and the scaled Solar System r-process-only abundances
of \citep{siqueira13}, adapted mostly from \cite{Simmerer.Sneden.ea:2004}. The stellar and Solar System distributions were normalized to agree for the element Eu (Z =  63), and vertical shifts were
then applied in each case for plotting clarity.
The stellar
abundance sets are (a) CS 22892-052 \cite{Sneden.Cowan:2003}, (b) HD 115444 \cite{Westin.Sneden.ea:2000}, (c) BD + 17
3248 \citep{Cowan.Sneden.ea:2002}, (d) CS 31082-001 \citep{siqueira13}, (e) HD 221170 \citep{Ivans.Simmerer.ea:2006}, (f) HD 1523 +0157 \citep{frebel07}, (g) CS 29491-069 and (h) HD 1219-0312 \citep{Hayek.Wiesendahl.ea:2009}(Hayek et al., 2009), (i) CS 22953-003 \citep{Francois.Depagne.ea:2007}, (j) HD 2252-4225 \citep{Mashonkina:2014}, (k) LAMOST J110901.22 +
075441.8 \citep{Li.Aoki.ea:2015}, (l) RAVE J203843.2-002333
\citep{Placco.Holmbeck.ea:2017}, and (m) 2MASS J09544277 + 5246414 \citep{Holmbeck.Beers.ea:2018}. Bottom panel: mean abundance differences for the
13 stars with respect to the Solar System r-process values. Figure reprinted with permission from \cite{Cowan.Sneden.ea:2021}.}
\label{fig:stars}
\end{figure}

Many observations of r-process-rich metal-poor Galactic halo stars indicated a relative solar system r-process abundance pattern \citep[see e.g.][]{gilroy88,sneden94,sneden96,Sneden.Cowan:2003,sneden03,Sneden.Cowan.Gallino:2008}.
These stars, by definition metal poor and having heavy element abundances smaller than the solar values, seemed to show element-to-element abundance ratios in the same proportions as in the solar system, i.e., a scaled solar system pattern. This is illustrated for a number of stars in 
Fig.\ref{fig:stars}, taken from \cite{Cowan.Sneden.ea:2021}. 
All of the stellar and solar system  abundance curves were normalized to the element Eu with vertical offsets for comparisons. It is seen that the same abundance pattern,  close to solar - with exceptions at the lightest and heaviest r-process elements - occurs in all of these stars. This suggests that the r-process responsible for these observations has operated in a similar manner across the Galaxy and provides constraints on the conditions for r-process nucleosynthesis resulting in this abundance pattern. However, it should be noticed that all these observations were taken from so-called r-II stars, i.e., r-process-rich stars (as indicated in the figure caption). 

The observed r-process abundances, which show dominant contributions at the metallicities of interest, have been divided by observers into subclasses of observed abundance patterns: limited-r, r-I, and r-II, which show different enhancements of Eu in comparison to Fe. Of the stable isotopes $^{151}$Eu and $^{153}$Eu, the latter one (amounting in the solar composition to 52\% of the element Eu) is dominated by the r-process \citep[see e.g.][]{Prantzos.Abia.ea:2020}. The different subcategories of Eu enhancement, introduced above, are determined by their [Eu/Fe] ratios: 
limited-r stars with [Sr/Ba]$>$0.5 and [Eu/Fe]$<$0.3, 
r-I stars with 0.3$<$[Eu/Fe]$<$0.7, and 
r-II stars with [Eu/Fe]$>$0.7 \citep[see e.g.][]{Holmbeck.Hansen.ea:2020,Saraf.ea:2023,Saraf.ea:2025,Mishenina.ea:2024}. [Eu/Fe]=0, i.e. the solar abundance ratio between these elements, which is in the range of limited-r stars, corresponds to a ratio of Eu/Fe$=1.15\times 10^{-7}$ \citep{Lodders.Bergemann.Palme:2025}. [Eu/Fe]=0.5, which is in the range of r-I stars, corresponds to an abundance ratio Eu/Fe=$3.6\times 10^{-7}$.
[Eu/Fe]=1.3, which is in the range of r-II stars, corresponds to an abundance ratio Eu/Fe=$2.3\times 10^{-6}$.

It is found that limited-r stars have a much steeper abundance curve as a function of mass number $A$, still containing elements as heavy as Eu but without or with vanishing amounts of third r-process peak elements. Whether r-I and r-II stars, while clearly showing differences in their [Eu/Fe] ratios, have similar abundance patterns for the heavy elements is still debated. \cite{Roederer.eauniversal:2022} and \cite{Racca.ea:2025} find a "universality" among the heavy elements (while differing in the ratio to Fe). On the other hand, a good fraction of r-II stars exhibit a so-called actinide boost \citep{Mashonkina:2014,Shah.ea:2026}. Utilizing correlation analyses for the element abundances \cite{Farouqi.ea:2022} and \cite{Farouqi.ea:2025} argue for different nucleosynthesis origins among r-I and r-II stars. 

\section{Interpreting the r-process impact in the Galaxy}\label{sec12}
In this section we aim to understand three aspects of the imprint of the individual possible r-process sites previously presented  in Section \ref{sec:sites}: (a) how important are the different contributions to explain the total solar r-process abundances, (b) can one link the observed [Eu/Fe] ratios of limited-r, r-I, and r-II stars to an origin from these individual sites, and (c) are the identifications based on (a) and (b) consistent with the time or metallicity ([Fe/H]) evolution found in observations of low-metallicity stars?

\subsection{Which events can explain the amount of solar r-process abundances?}

\begin{figure}[h!]
    \centering
    \includegraphics[width=0.9\linewidth]{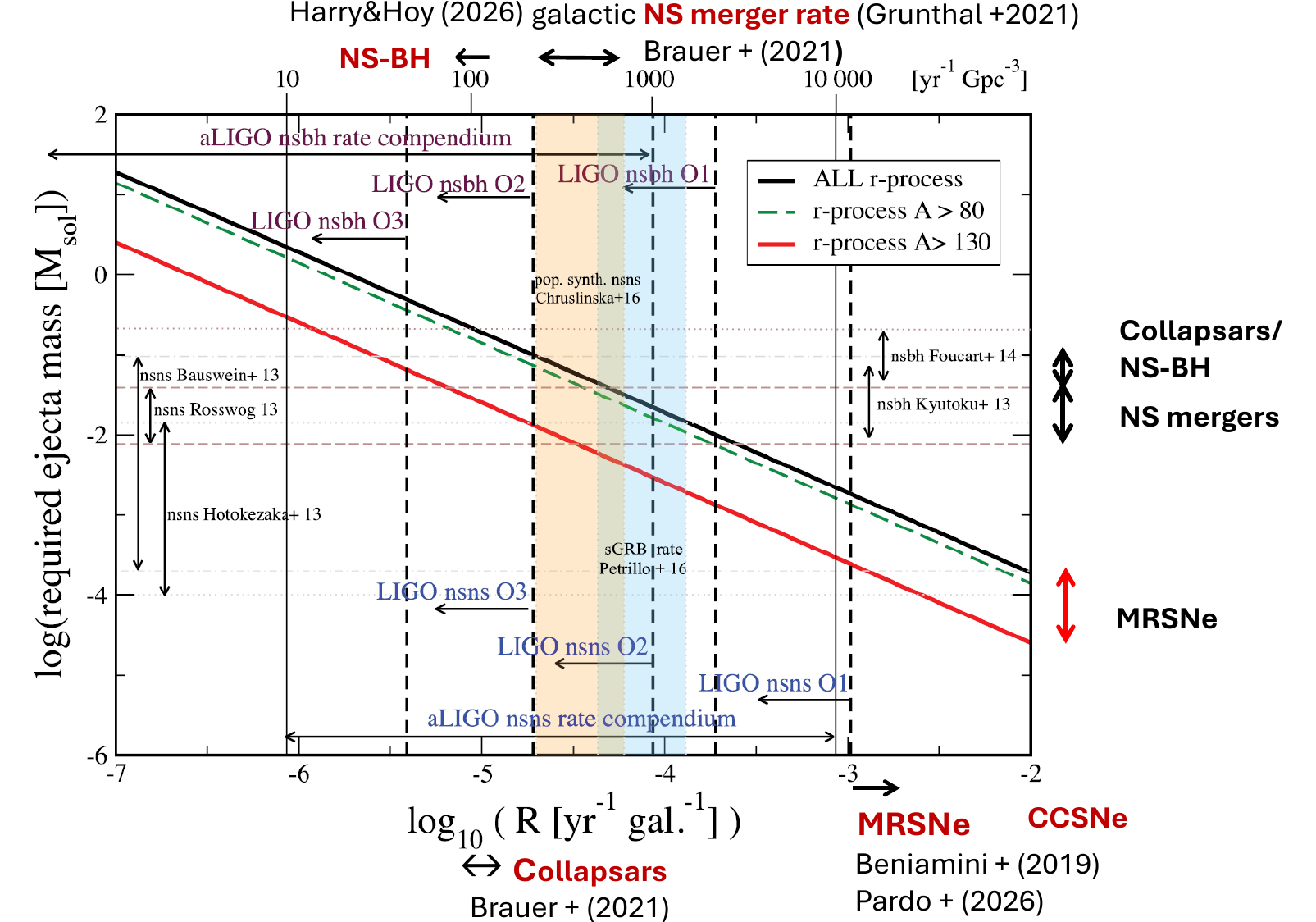}
    \caption{Required occurrence frequencies in units per year and galaxy (or yr$^{-1}$Gpc$^{-3}$) in relation to the necessary amount of r-process ejecta for reproducing the solar r-process abundances; adapted from \cite{Rosswog.Feindt.ea:2017}. For updates see Fig.5 of \cite{Rosswog.Korobkin:2024}, which includes also the double neutron star merger rate in the Milky Way determined by \cite{Grunthal.Kramer.ea:2021} to be $3.2 ^{+1.9}_{-0.9}\times 10^{-5}$ per year, covering the yellow shaded region of this graph. Further updates are based on the few gravitational wave observations identified with neutron star mergers and neutron star - black hole mergers \citep{Abac.ea:2025,Fishbach.ea:2026}, which cover a large uncertainty range extending to lower numbers, see the discussion in Sections \ref{sec:nsm} and \ref{sec:other}. For the other entries in these graphs see the references in both Rosswog papers, but also the arrows (with references) on the abscissa and the predicted amounts for the different sites given on the right ordinate (according to the discussions in the previous section;  see especially the red "arrow bar" for magneto-rotational supernovae). Figure adapted from \cite{Rosswog.Feindt.ea:2017} with additional entries provided on the abscissa and ordinate.} 
    \label{fig:rosswog17}
\end{figure}

If the solar r-process abundance pattern would be due to one dominant site, the basic requirements for such a site are presented in Fig.\ref{fig:rosswog17} in terms of necessary ejecta amounts vs. the occurrence frequency in the Galaxy, completely independent of the nature  of the astrophysical site. We examine here the list of sites discussed in Section \ref{sec:sites} and test their possible importance with the aid of the following figure.

\begin{itemize}
    \item 
If heavy r-process elements of the solar composition would come from regular supernovae, this would require (in each supernova explosion) the ejection of about $10^{-4}$M$_\odot$ integrated over all r-process nuclei, assuming a continuous production with supernova occurrence rates of approximately  1/100yr in the Galaxy. As seen in our discussion in section \ref{sec:regsne}, it is clear that this requirement is not fulfilled.
\item
For magneto-rotational supernovae (see section \ref{sec:MHD}) we quote a number of $5\times 10^{-7}-3\times 10^{-6}$M$_\odot$ ejecta of Eu. If scaled, consistent with a solar r-process pattern, this would correspond to a total mass of r-process ejecta of $5\times 10^{-4}-3\times 10^{-3}$M$_\odot$. With a rate of magneto-rotational supernovae amounting to about 1/10 of all CCSNe (see the discussion in Section \ref{sec:MHD}), this would vary from a substantial contribution to even an overproduction in the relation shown in Fig.\ref{fig:rosswog17} (see the red error bar). However, we know that the average magneto-rotational supernova does not produce the heaviest r-process nuclei, leading only to a weak or limited r-process pattern, producing some Eu. Knowing the total ejected mass of nuclei with $A>130$, we scale the given Eu masses with the ratio of A($>$130) vs. Eu masses, according to the relatively flat abundance behavior in the mass range above the 130 peak, as seen in Fig.\ref{fig:Nishimura172}. Thus, multiplying the Eu mass by about 60, leads to a total mass for $A>130$ in the range of $3\times 10^{-5}$ to $1.8\times 10^{-4}$M$_\odot$. Comparing this range with the values on the red line (standing for the solar r-process beyond $A=130$) at occurrence frequencies around $10^{-3}$yr$^{-1}$gal$^{-1}$ (about 1/10 of the CCSN rate), indicates a variation from a 10\% contribution to the solar r-composition beyond 130 to a 100\% contribution by ejecta from MRSNe alone. The latter value is clearly too large, knowing the strong decline beyond $A=130$ in the predicted abundance curve of a weak r-process. Since MRSNe were the only (among the previously discussed) sites with a measurable co-production of Fe and Eu, leading automatically to a correlation between these elements in the abundances of a newly formed star with such a contribution, we can link MRSNe to observations of limited-r stars. When looking at Fig.4 in \cite{Cowan.Sneden.ea:2021}, we find that the two limited r-stars HD 88609 and 122563 in that plot have a typical underabundance in comparison to solar of a factor 10 in the heavy mass region. Therefore, we come to the conclusion that one should rather favor the low value of $5\times 10^{-7}$M$_\odot$ of Eu for the typical MRSN contribution to make it compatible with observations in limited-r stars. Thus, we will reduce the MRSN input for further considerations to the above given lower limit. This still keeps the correlation between Fe and Eu for this environment, as found in low-metallicity limited-r stars \citep[see the next subsection and][]{Farouqi.ea:2025}, but this source is not a dominant r-process contributor for the heavier elements beyond $A$=130.
Magnetar giant flares (taking place after magnetar formation in magneto-rotational supernovae) with a total amount of $10^{-6}$M$_\odot$ of r-process nuclei would contribute about 0.1\% of r-process matter to the earlier supernova. If, however, multiple such flares would occur after each magneto-rotational supernova, this percentage would be as high as 1\%. If the magnetar formation rate is higher than the MRSN rate, this fraction could even increase further.

\item
In section \ref{sec:colhyp} we refer to \cite{Siegel.Barnes.Metzger:2019}, who predict $>10^{-1}$M$_\odot$ of r-process matter in collapsars/hypernovae. \cite{Brauer.ea:2021} predict a slighly smaller amount of $7\times 10^{-2}$M$_\odot$. If these would qualify as a dominant contribution to the solar r-process, this would require a frequency of about $3\times 10^{-5}$ per year and galaxy, or about one event per 300 CCSNe (of which 1-2 per 100yr take place in the Galaxy), according to Fig.\ref{fig:rosswog17}. This is larger by about a factor of 3 than the estimates provided by \cite{Brauer.ea:2021} for collapsar occurrence rates. Taking this at face value, would indicate that collapsars can provide a substantial contribution, but they would not be the dominant r-process site.  

\item
In section \ref{sec:nsm} we quote an amount of total r-process ejecta of $(1-5)\times 10^{-2}$M$_\odot$ for neutron star mergers. This number is smaller by a factor of 2-10 than the one for collapsars/hypernovae. Thus, if neutron star mergers would be the dominant r-process production site, this would require about 1 event per 30-200 CCSNe, making it marginally consistent with estimates by \cite{Grunthal.Kramer.ea:2021} and \cite{Brauer.ea:2021} in Fig.\ref{fig:rosswog17}. This would indicate that neutron star mergers are the dominant r-process site, overlapping with the numbers given by \cite{Chen.Li.ea:2024} of $320^{+490}_{240}$Gpc$^{-3}$yr$^{-1}$ or $10^{-5}$-10$^{-4}$ per year and galaxy, but see also discussions related to the recent Gravitational Wave Transient Catalogue \citep[GWTC-4 entries,][]{Abac.ea:2025,Fishbach.ea:2026} which extend to lower rates.
\item
In section \ref{sec:other} we quote the r-process ejecta mass for neutron star - black hole mergers by \cite{Wanajo.Fujibayashi.ea:2024} to be of the order $(3-7)\times 10^{-2}$M$_\odot$, therefore being on average about a factor of 2 or more larger than in neutron star mergers. If these mergers occur less frequently than neutron star mergers by that ratio, they could also be a significant r-process site. This is indicated in Fig.\ref{fig:rosswog17}, allowing it to be dominant for actinide boosts. 
Another source mentioned in that section is due to the disruption of subminimal-mass neutron stars which can occur during mass transfer in compact binaries from the less massive neutron star. Such evolution is most likely in compact binaries with large mass ratios, therefore neutron star - black hole binaries have the largest probability to lead to such events \citep{Martineau.ea:2026}. The authors \citep{Yip.Leung.ea:2023,Ignatovsky.ea:2023} state that comparable or even higher r-process ejecta masses than for neutron star mergers are expected. If this were to occur in each neutron star - black hole merger event, this would bring up the total combined amount of r-process ejecta similar to the above mentioned collapsar/hypernova contribution.
\end{itemize}

We had a relatively long discussion of magneto-rotational supernovae in this subsection, but it helped to link their weak r-process feature, combined with a non-negligible Fe ejection, to observations of limited r-stars. Furthermore, this subsection led also to the result that neutron star mergers could be the dominant r-process site in the Galaxy, while collapsars/hypernovae and neutron star - black hole mergers can provide substantial, but not dominant contributions, except possibly in the case of actinide boost stars.

\subsection{Evolution as a function of metallicity}

While in the previous subsection we concentrated on the possible overall contributions of the discussed astrophysical sites, we want to focus in this subsection on the time or metallicity evolution found for r-process elements, as well as how the different subclasses (limited-r, r-I, and r-II stars) could be linked to the previously discussed sites. As was outlined in great detail in Section \ref{sec:sites}, magneto-rotational supernovae, neutron star mergers, and collapsars/hypernovae, as well as neutron star - black hole mergers follow  in this sequence with increasing r-process yields. As a working hypothesis we want to test whether this could be consistent with linking these sites with the three subclasses of r-process enrichment in low-metallicity stars. But before that we want to give a short overview of  their appearance in Galactic evolution.

\begin{figure}[h!]
    \centering
    \includegraphics[width=1.05\linewidth]{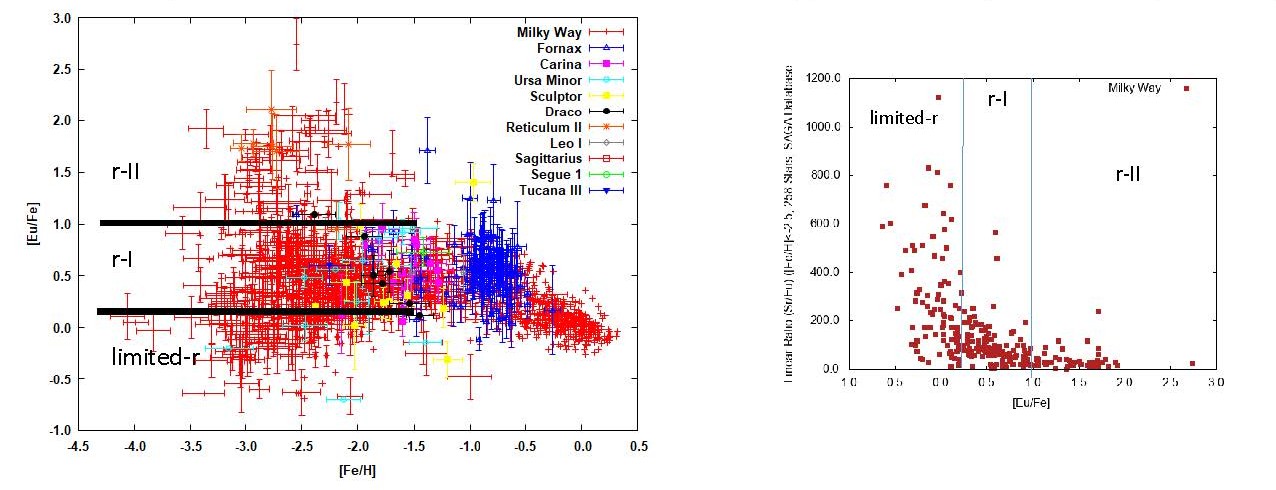}
    \caption{Left: [Eu/Fe] ratios observed in a large number of low-metallicity stars as a function of [Fe/H]. For metallicities [Fe/H]$<$-2 the scatter can be as high as a factor of 100. Right: Sr/Eu as a funtion of [Eu/Fe]. Limited-r stars show a factor of 10 higher Sr/Eu ratios than r-I and r-II stars, i.e. a much steeper abundance decline as a function of nuclear mass number. The data are based on the SAGA \citep{Sagadatabase} and JINA \citep{JINAbase:2018} databases; figures adapted from \cite{Farouqi.ea:2022}.}
    \label{fig:EuFeSr}
\end{figure}

In Fig.\ref{fig:EuFeSr} we show [Eu/Fe] ratios as a function of metallicity and the (Sr/Fe) ratios as a function of [Eu/Fe]. [Eu/Fe] shows a large scatter of approximately a factor of 100 for metallicities below [Fe/H]$\approx$-2. This is the time/metallicity in Galactic evolution when the ejecta contributions of individual explosive events are still visible in the composition of new-born stars. Apparently at metallicities near -2 enough overlapping events have contributed, demonstrating that from then on the composition of new-born stars exhibits an average over all contributing sources and the scatter becomes small. This average also occurs after the decline at about [Fe/H]=-1, the time in Galactic evolution when type Ia supernovae enter with their larger Fe ejecta. The timing is delayed both for white dwarf formation in Galactic evolution and for the mass transfer in a binary system, which leads to the explosion of the white dwarf. This average behavior is well explained in classical homogeneous "chemical evolution" models with the instantaneous mixing approximation IMA, described e.g. in \cite{matteucci86,Timmes.Woosley.Weaver:1995,Matteucci:2012,Matteucci:2021} and with relation to the average [Eu/Fe] as well \citep[see e.g.][]{Matteucci.Romano.Arcones.ea:2014,Cote.Belczynski.ea:2017,Cote.Fryer.ea:2018,Molero.ea:2025}. But, even in this classical approach, there exist indications that neutron star mergers might not be the only necessary r-process production site \citep{Molero.ea:2025}.  

Inhomogeneous chemical evolution approaches, not accounting for a spatial averaging and averaging over a number of individual stellar sites,  which contribute to the interstellar medium from which new stars are born \citep{wehmeyer15,Thielemann.Eichler.ea:2017,Cote.Eichler.ea:2019,Cote.Yague.ea:2019,Wehmeyer.ea:2019,Siegel:2019,Siegel:2022,VandeVoort.ea:2020,VandeVoort.ea:2022,Kobayashi.Mandel.ea:2023}, lead to the same conclusion. This apparently requires the need for at least one additional r-process astrophysical site responsible for limited-r stars. Whether other sites (discussed in the previous sections) - in addition to neutron star mergers - are also required, will be the focus when examining the early Galactic evolution below metallicities of -2, where the observed abundance patterns are still affected by individual events. 
It can be seen in Fig.\ref{fig:EuFeSr} that limited-r stars typically have very high Sr/Fe ratios, i.e. their heavy element abundances are tilted towards the low-mass end, and third r-process peak elements are not detected in these limited-r stars. At these low Galactic metallicities, stellar surface abundance patterns indicate the composition of the interstellar medium out of which the star formed, and only one or few nearby stellar explosions contributed to that composition.  This therefore points to the astrophysical explosion site from which this abundance pattern originated. Following our earlier discussion on astrophysical sites, the pattern of limited-r stars suggests a weak r-process as predicted from magneto-rotational supernovae (and possibly their subsequent magnetar giant flares). The correlation analysis of \cite{Farouqi.ea:2025}, which shows overall a mild (but existing) correlation of weak r-process elements with Fe, and especially a strong correlation in limited-r stars, underlines this argument. The question still remains whether the patterns of r-I and r-II stars originate from different astrophysical sites. A statistical cluster analysis \citep[see Fig.4 of][]{Farouqi.ea:2022} seems to support the division into limited-r, r-I, and r-II stars, introduced previously by observers. Of the possible sites discussed in the preceding sections, two of them still remain in addition to magneto-rotational supernovae (producing a limited-r pattern): collapsars/hypernovae and compact binary mergers (consisting of neutron star mergers and neutron star - black hole mergers). Would we expect a different r-process abundance pattern from them? Among the observations, the Eu/Fe ratio is different for r-I and r-II stars, but the overall abundance pattern among heavy elements seems very similar \citep{Roederer.eauniversal:2022,Racca.ea:2025}.
However, abundance patterns with a so-called actinide boost (indicating $Y_e$ conditions as shown in Figs.\ref{fig:meng-ru} and \ref{fig:rabund-Aye}) as predicted e.g., for neutron star - black hole mergers \citep{Wanajo.Fujibayashi.ea:2024} are seen in a fraction of r-II stars \citep{Mashonkina:2014,Holmbeck.Beers.ea:2018,Shah.ea:2026}.

\begin{figure}[h!]
    \centering
    \includegraphics[width=0.7\linewidth]{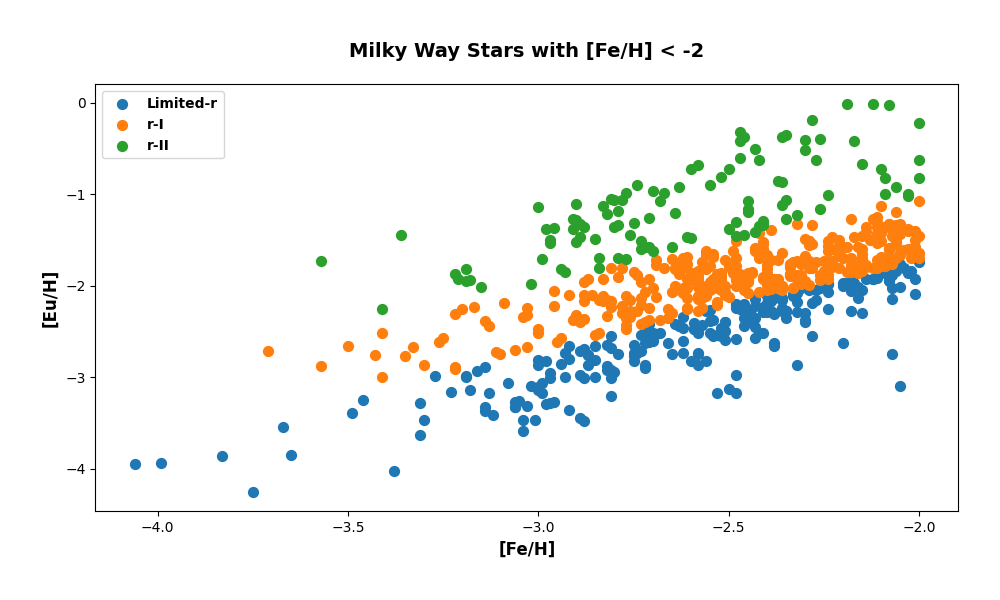}
    \caption{[Eu/Fe] abundances as a function of [Fe/H] for low-metallicity limited-r (blue), r-I (orange),
and r-II (green) stars with [Fe/H] $<$ -2. An apparent change in the onset of the three components can be noticed at different metallicities; from \cite{Farouqi.ea:2025}, copyright by authors.}
    \label{fig:farouqFeH}
\end{figure}

One can now ask the question whether these different categories of stars with heavy elements associated with the r-process emerged at different times during the Galactic chemical evolution. This would be reflected in their metallicity distributions.   
Fig.\ref{fig:farouqFeH} displays stars with available Eu and Fe abundances in the [Eu/Fe] vs. [Fe/H] plane. The onset of the three categories indicates clearly that the limited-r pattern appears in the most metal-poor (earliest?) extremely metal-poor (EMP) stars for metallicities below [Fe/H] $\approx-$4. The r-I and r-II stars follow at [Fe/H] $\approx-$3.5, 
which could be compared with the expected frequency of events. If we associate limited-r stars with the contributions from magneto-rotational supernovae, i.e. supernovae that produce neutron stars with high magnetic fields (magnetars), this accounts for about 10\% (or more?, see Section \ref{sec:MHD}) of regular CCSNe \citep[e.g.][]{Beniamini.Hotokezaka.Horst.ea:2019,Pardo.Rea.ea:2026}.
The lowest-metallicity stars have formed from gas enriched by yields of the first CCSNe. Although the supernova remnant of such an explosion is expected to exhibit an [Fe/H] ratio of about -3 \citep{Ryan.ea:1996}, new-born stars, forming from the ISM and polluted by only about 1\% of such an explosion, would then show [Fe/H] $\approx-$5. This is commensurate with the findings for EMP stars which can have even smaller values \citep[e.g.][]{Norris.Christlieb.ea:2007,Norris.Christlieb.ea:2012,Frebel.Norris:2015,Nordlander.ea:2019}. For events which occur with 1/10 of the frequency of regular CCSNe (like magneto-rotational supernovae) one might then expect them to emerge first at metallicities [Fe/H] $<-$4, consistent with the findings for limited-r stars in Fig.\ref{fig:farouqFeH}. 

Nucleosynthesis events that take place only after about 100 or more regular CCSNe enriched the ISM, can be expected to lead to stars with higher metallicities ([Fe/H]$\approx$-3 and above). It was argued in \cite{Farouqi.ea:2022} that collapsars as well as compact binary mergers have a frequency lower than regular CCSNe by approximately a factor of 120. (We will discuss such numbers further in the following subsubsections.) These order of magnitude estimates appear to be reflected in what is shown in Fig.\ref{fig:farouqFeH} for the different r-process-enhanced stars. Although from that figure alone, one cannot argue that the three categories limited-r, r-I, and r-II stars represent the enrichment scenarios of star-forming regions by different contributing explosion sites, we note that correlation analysis (comparing Pearson and Spearman correlation values for Fe and Eu) would, however, support this conclusion \citep{Farouqi.ea:2025}.

\subsubsection{How to reproduce limited-r, r-I, and r-II [Eu/Fe] ratios based on ejecta from various r-process sites?}
Here we want to quantify the statements from the previous discussions by basing them upon the predicted r-process production in comparison to observed limited-r, r-I, and r-II stars. We should caution here that this is somewhat of a bold move, especially as the abundance predictions utilized here bear a large degree of uncertainty and the three categories of low-metallicity stars also cover extended and almost continuous ranges in [Eu/Fe], rather than showing only typical values. However, among other reasons a statistical cluster analysis \citep[see Fig.4 of][]{Farouqi.ea:2022} apparently supports the division into limited-r, r-I, and r-II stars, introduced previously by observers. The main reason for this approach comes from the increasing r-process (including Eu) production from MRSNe (which also come with a non-negligible amount of Fe) to neutron star mergers and finally to collapsars/hypernovae, as well as to neutron star-black hole mergers. If these sites are also characterized by decreasing occurrence frequencies, their imprint would be accompanied by a sequence of an increasing underlying Fe floor from regular CCSNe, as a function of Galactic evolution. In the following discussion we want to verify whether our working hypothesis to explain this behavior from the contributions of such different sources makes sense.

The typical value of [Eu/Fe]$\approx$0 for limited r-stars corresponds to an abundance ratio Eu/Fe=$1.15\times 10^{-7}$ \citep{Lodders.Bergemann.Palme:2025}. We consider/suggest here that magneto-rotational supernovae cause the abundance imprint in limited-r stars.  Taking the predicted ejecta composition with $5\times 10^{-7}-3\times 10^{-6}$M$_\odot$ of Eu and about $5\times 10^{-2}$M$_\odot$ of Fe from section \ref{sec:MHD},  leads to a mass ratio of $10^{-5}-6\times 10^{-5}$ or to an abundance ratio of $3.6\times 10^{-6}-2.2\times 10^{-5}$, when considering mass numbers of 153 and 56. It should be kept in mind that this large spread in abundance predictions includes the whole range of the i-models in \cite{Nishimura.Sawai.ea:2017}. If we assume  that the average of all MRSNe models lies close to the minimum in this range, i.e. at $3.6\times 10^{-6}$ (supported by the discussion in the previous subsection in relation to Fig.\ref{fig:rosswog17}), this is still higher by a factor of 31 than the average abundance ratio found in limited-r stars, which corresponds to [Eu/Fe]=0. This can be explained if the Fe ejected in this event ($\approx0.05$M$_\odot$) adds to an already existing floor of Fe produced by preceding "regular" CCSNe (typically producing about 0.1M$_\odot$ of Fe) occurring with a higher frequency. Thus, in order to obtain the typical abundance ratio of limited-r stars, the mass in Fe has to be enhanced by this factor $(31 \cdot 0.05  = 1.55 =(0.05 + n \cdot 0.1)$, resulting in $n=15$. When taking the upper limit from limited r-star observations [Eu/Fe]=0.3, this would reduce 31 by a factor of 2 down to 15.5, leading to $15.5\cdot 0.05  = 0.78 =(0.05 + n \cdot 0.1)$ and an $n\approx7$. Thus, regular CCSNe per MRSN ratios of roughly 7-15 would result (where 7 corresponds to the higher limit [Eu/Fe]=0.3 for limited-r stars), which is approximately consistent with the previous considerations related to Fig.{\ref{fig:rosswog17}}.

If we want to make neutron star mergers, with a higher ejected mass of r-process matter, responsible for r-I stars with an average [Eu/Fe]=0.5, we have to aim for an Eu/Fe abundance ratio of $10^{0.5}\cdot 1.15\times 10^{-7}\approx3.6\times 10^{-7}$ (see section \ref{sec:earlygal}) or a Eu/Fe mass ratio of about $9.8\times 10^{-7}$. In this case the typical predicted Eu mass is about $2\times 10^{-5}$M$_\odot$ (see \ref{sec:nsm}, however also note that \cite{Cote.Fryer.ea:2018} quote values which could be smaller). To attain the Eu/Fe mass ratio in r-I stars would then require an Fe mass of about 20M$_\odot$ to the Eu ejecta in the corresponding explosions, which is impossible to produce in a neutron star mergers. In principle, a neutron star merger ejects negligible amounts of Fe, but it is possible (although kick velocities might reduce this effect) that the Fe amount of the two preceding supernovae (0.2M$_\odot$) might still be contained in the extended ejecta of the merger. In addition - at that time of Galactic evolution - a floor of Fe by produced by $n$ preceding supernovae would be contained in the local ISM. Thus, this would mean that 20=(0.2+$n\cdot$ 0.1), requires $n\approx 198$, i.e. the Eu would have to be dispersed among the Fe of almost 200 preceding CCSNe. This is consistent with the  ratio of CCSN explosions per neutron star merger, as discussed in the preceding subsection. It should be noted that, if the upper limit [Eu/Fe]=0.7 is used - rather than the typical 0.5 - for r-I stars, this would reduce the required Fe floor by a factor of about 1.6 and thus the number of preceding regular CCSNe by the same amount, i.e. $n\approx 125$.

Collapsars/hypernovae, as well as neutron star - black hole mergers, were predicted in sections \ref{sec:colhyp} and \ref{sec:other} to produce 3 to 5 times the amount of r-process matter as neutron star mergers, including Eu. These higher amounts of Eu ejecta could make them candidates for r-II stars with [Eu/Fe] ranging from 0.7 to more than 2. We use here an average value of 1.3. This corresponds to an Eu/Fe abundance ratio of $10^{1.3}\cdot 1.15\times 10^{-7}\approx 2.3\times 10^{-6}$ (see section \ref{sec:earlygal}) or a mass ratio of approximately $6.3\times 10^{-6}$. With the typical fraction of the Eu mass being approximately $10^{-3}$ of the entire r-process matter in a solar composition, this translates the above mentioned r-process production to Eu masses of $7\times 10^{-5}-10^{-4}$M$_\odot$. To be consistent with a value of [Eu/Fe]=1.3, would then require an Fe mass of approximately 11-16M$_\odot$. This would be impossible for one such event, which typically produces only about 0.5M$_\odot$ in the case of collapsars, and even less for neutron star - black hole mergers.
Thus, this would mean that 11 to 16=(0.5+$n\cdot$ 0.1), requiring $n\approx 105-155$, i.e. the Eu would be dispersed among the Fe of about 100-150 preceding CCSNe. The case would be similar for collapsars/hypernovae and neutron star - black hole mergers.
This derived number is different than the limits given in the previous section and in Fig.\ref{fig:rosswog17}, where - on average - one event for about 1000 CCSNe is indicated. However, the present result was obtained assuming a typical [Eu/Fe]=1.3. Raising this value to [Eu/Fe]=2, still within the r-II range,  would enhance the ratio by a factor of 5, i.e. up to $n=750$, which brings it close to the previously discussed values.

The frequencies determined in this manner for neutron star mergers and collapsars/hypernovae or neutron star - black hole mergers are, therefore, comparable to the ones discussed in reference to Fig.\ref{fig:farouqFeH}, when considering the working hypothesis in the first paragraph of this subsection. It should be noted that the derivation of these numbers was based on taking at face value the production masses of the model simulations, which clearly have to be taken with large uncertainties. In addition, the typical [Eu/Fe] ratios for limited-r, r-I, and r-II stars utilized here are taken from an almost continuous spectrum seen e.g. in the left part of Fig.\ref{fig:EuFeSr} or also in Fig.3 of \cite{Saraf.ea:2023} and Figs.6 and 8 of \cite{Saraf.ea:2025}. The present discussion leads to the conclusion that the occurrence frequencies of different types of events, indicated in the previous subsection and the present subsubsection, are consistent with reproducing the total amount of r-process ejecta, as well as demonstrating specific abundance patterns at related metallicities in Galactic evolution. This in turn implies a required floor of Fe ejecta from the preceding CCSNe. We still have to verify whether this is also consistent with the timing of events in Galactic evolution models. The main question is at what metallicity does r-process Eu from the rarer events (neutron star mergers and collapsars or neutron star-black hole mergers) appear. 

It should also be noted why we suggested neutron star mergers to be responsible for r-I stars and collapsars/hypernovae,  and/or neutron star - black hole mergers to be responsible for r-II star patterns, showing in general similar distributions solely among the heavy r-process elements. One obvious reason would be that the latter produce up to a factor of 5 times larger amounts of Eu, leading to higher [Eu/Fe] ratios. Another point is that a good fraction of r-II stars show actinide boost abundance patterns, apparently predicted by neutron star - black hole simulations \citep{Wanajo.Fujibayashi.ea:2024}. 

\subsubsection{How to reproduce the r-process input as a function of metallicity [Fe/H] in galactic evolution models?}
Many inhomogeneous galactic evolution simulations (allowing for individual local events, which can result in spatially inhomogeneous abundance distributions) had r-process contributions from neutron star mergers appear only at a metallicity of about [Fe/H] = -2.5 \citep[e.g.][]{wehmeyer15,thielemann17a} - see also the olive-dotted lines in Fig.32 of \cite{Kobayashi.Karakas.Lugaro:2020} and the corresponding yellow line in Fig.3. of \cite{Grichener:2025}. Those calculations utilized a neutron star merger rate of $10^{-5}$ per year and galaxy, which could actually be larger by a factor of 10 \citep[as stated in][]{Kobayashi.Karakas.Lugaro:2020}. This would move the curve upward by that factor and have it agree with observations at [Fe/H]=-1, but this still leads to a too steep decline towards lower metallicities. That latter part depends on the timing requirement that (i) two core-collapse supernovae have to explode in a binary system (fast), (ii) the binary system has to survive this explosion, and (iii) in addition an inspiral has to take place due to energy loss via gravitational wave emission, leading finally to the merger event. Massive binary  stellar systems on this evolutionary path result in a delay-time distribution \citep[DTD, see e.g.][]{Maoz.Nakar:2025}, which measures the time between star formation and the neutron star merger. 

\cite{Grichener.Kobayashi.Soker:2022} showed that with a delay time of 10-20Myr the steep decline towards low metallicities, mentioned above, could be avoided for the average [Eu/Fe] as a function of [Fe/H]. They attributed this  to the occurrence of common envelope jet supernovae, originating from mergers of neutron stars with the cores of massive stars, discussed previously in section \ref{sec:other}. They could occur earlier and more frequently than neutron star mergers. However, this scenario - until now only investigated in analytical approaches - is still waiting for multi-dimensional simulations of such events, which would confirm the existence of a high density in the inner layers of the accretion disk (with high electron Fermi energies to cause neutron-rich matter by electron captures) and the possible mixing of neutron star matter into the ejected inner disk matter \citep{Soker:2025}. Such conditions are required to produce a full r-process up to the heaviest nuclei.

The delay-time distribution DTD of gravitational-wave-driven neutron star mergers is predicted to follow a power law,  proportional to $t^{-1}$. However, \cite{Maoz.Nakar:2025} found that the observed distribution of mergers is well reproduced by a two-population model, combining mergers born with a $t^{-1}$ DTD and a second population born with a $t^{-2}$ DTD. Although the vast majority of all mergers take place within 1Gyr, a substantial excess between 10 and 100 Myr is found in comparison to a pure $t^{-1}$ power law. Such changes would reproduce the observed [Eu/Fe] as a function of [Fe/H] for neutron star mergers, as discussed above. The argument by \cite{Beniamini.Piran:2019} that the kick from the second supernovae explosion can result in ultra-fast mergers would also support the claim that neutron star mergers can account for the r-process in the early galaxy.

\cite{VandeVoort.ea:2020,VandeVoort.ea:2022} found an "optimized DTD", allowing for r-process contributions from mergers to show up as early as [Fe/H]$<$-3 in Galactic evolution, if the merger rate is about 0.3\% of the CCSN rate (see their Fig.3). Based on Fig.\ref{fig:rosswog17} of \cite{Rosswog.Feindt.ea:2017} and Fig.4 of \cite{Chen.Li.ea:2024} this leads to the conclusion that a rate on the order of 0.3\% of the CCSNe rate is required. This corresponds to about a rate of $R=$500-600Gpc$^{-3}$yr$^{-1}$, in comparison to the 2021 limits given by \cite{Abbott.eaX:2021} of 310$^{+490}_{-240}$Gpc$^{-3}$y$^{-1}$. \cite{Kobayashi.Mandel.ea:2023} (see their Fig.4) could show that with a metallicity-dependent DTD neutron star mergers can reproduce the observations of [Eu/Fe] vs. [Fe/H].
With these considerations, it seems that neutron star mergers - happening at a frequency once per a few 100 CCSNe - can explain the appearance of r-process elements like Eu at metallicities as low as [Fe/H]$<$-3, but it should be noted that the lower part of the recently determined error range of 7.6- 250 Gpc$^{-3}$yr$^{-1}$ from the Gravitatioal Wave Transient Catalog (GWTC-4)  \citep{Abac.ea:2025} could create a tension for this interpretation \citep[see also the discussion in][]{Fishbach.ea:2026}.

It remains to be seen whether (after being confirmed by self-consistent multi-D simulations) the suggested common-envelope-jet supernovae, resulting from a merger of a neutron star with the core of a massive star \citep{Grichener.Kobayashi.Soker:2022,Jin.Soker:2024,Grichener:2025}, can play a comparable role as well.

\subsection{Earliest and Latest Imprints}

\subsubsection{Earliest Imprints and Cosmochronology}

Using the predicted r-process abundance production rates  among actinide nuclei, imprinted in the next generation of stars and decaying since that initial time, allows a determination of the decay time until the present observational element/isotope ratios were attained. This is mostly based on the ratio of $^{238}$U and $^{232}$Th,  which decay with half-lives of $4.5\times 10^9$y and $1.4\times 10^{10}$y, respectively,  eventually to the stable isotopes $^{206}$Pb and $^{208}$Pb. Simple chemical evolution models have been proposed for these investigations \citep{Fowler.Hoyle:1960,Schramm.Wasserburg:1970}. For early reviews see e.g. \cite{Cowan.Thielemann.Truran:1991} and \cite{Arnould.Goriely:2001}.
Later work focused on the individual decay times since the formation of stars in the early Galaxy 
\citep[e.g.,][]{Cowan.Pfeiffer.ea:1999,Schatz.ea:2002,Otsuki.ea:2003,Frebel.Kratz:2009,Roederer.Kratz.ea:2009,Huang.ea:2025}. This resulted in age determinations of these very old stars on the order of 13 billion years with limits from 11 to 14 Gyr. Assuming that these abundances are due to imprints from r-process contributions in the very early Galaxy, this provides also limits for the age of the Galaxy and the Universe.

\subsubsection{Latest Imprints from Nearby Events}
A completely different approach involves determining the latest near-by events which actually "polluted" our terrestrial environment. Two radioactivities $^{60}$Fe ($\tau_{1/2}=2.6\times 10^6$y) and $^{244}$Pu ($\tau_{1/2}=8.06\times 10^7$y) have been found in deep-sea sediments, both with lifetimes shorter than the time since the formation of the solar system. These deep-sea sediments grow their layers in time, and the composition of layers as a function of radius permits us to look into the history of radioactive contributions which penetrated from outside the earth. $^{60}$Fe is produced during the evolution of massive stars and ejected during the concluding supernova event. $^{244}$Pu is produced in a full (robust) r-process. \cite{Wallner:2016} reported the detection of live $^{244}$Pu deposits, archived in deep-sea sediments during the last 2.5$\times 10^6$y. If heavy r-process elements of a solar composition would come from regular supernovae, this would require (in each supernova explosion) the ejection of $10^{-4}$-$10^{-5}$ M$_\odot$ of
r-process nuclei, assuming a continuous production with supernova occurrence rates on the order 1/100yr (see Fig.\ref{fig:rosswog17}). 

Such a continuous production in the Galaxy with intervals much shorter than the $^{244}$Pu decay time, would lead to deposition rates which are inconsistent with the detections. The observed abundances were lower than expected by about 2 orders of magnitude.
\cite{hotokezaka15} have shown that this can be explained with the rarity of (compact merger?) events producing $^{244}$Pu, allowing for partial decay before incorporation on earth and leading to a varying time structure in contributions to the solar system, which is broadly consistent with the $^{244}$Pu abundances of the early Solar system material \citep{Turner.ea:2007}. Rare events with enhanced ejecta masses (beyond the ones required for supernova occurrence frequencies) could also explain solar abundances, with the option that the last archived event  
took place in a more distant past and Pu has decayed since then.

A more recent investigation by \cite{Wallner.ea:2021}, analyzing Pacific Ocean-crust samples, found two recent supernova contributions with $^{60}$Fe deposits originating about 2.5$\times 10^6$yr and 6.3$\times 10^6$yr ago. The somewhat surprising result is that both events seem to be accompanied by $^{244}$Pu as well, which - in principle - is not co-produced with $^{60}$Fe in supernovae, neither in regular core-collapse supernovae nor magneto-rotational supernovae, but possibly in collapsars/hypernovae (see Sections \ref{sec:regsne}, \ref{sec:MHD}, and \ref{sec:colhyp}). The interpretation is still unclear, ranging from a possible collapsar/hypernova contribution which would come with $^{60}$Fe from the massive star origin and also produce a full r-process. Other options are that the supernova ejecting the $^{60}$Fe also swept up interstellar medium dust condensates, which contain $^{244}$Pu \citep{Wehmeyer.Lopez.ea:2023,Wehmeyer.Lopez.ea:2024,Wehmeyer.Lopez.Cote:2025}. \cite{Wang.Clark.Fields:2023} suggest examining such options with lunar regolith samples.

\subsubsection{Related Meteoritic Inclusions}
Related to the previous discussion of the most recent radioactivities found in deep-sea sediments is also the question of how they were incorporated into dust condensates resulting from explosive origins and later found in meteoritic inclusions. Some of these processes, including among others the production of isotopes $^{129}$I, $^{244}$Pu, and $^{247}$Cm, have been discussed in a number of investigations \citep[see e.g.][]{Cote.ea:2021,Lugaro.ea:2022,Lugaro.Lopez.ea:2022}.
Also \cite{Nakashima.ea:2021} and \cite{Nakashima.ea:2025} have looked for initial $^{244}$Pu/$^{238}$U ratios in Ca-Al-rich meteoritic inclusions, using noble gas and trace element abundances, which represent the initial $^{244}$Pu/$^{238}$U ratio present in the solar system.

In a related study \cite{Meyer:2005} notes that the early solar system abundances  of $^{107}$Pd, $^{129}$I, and $^{182}$Hf could potentially be accounted for in an n-process-type environment related to explosive He-burning in a nearby massive star during its supernova explosion. This would be similar to what has been discussed in Sections \ref{sec:alphanp} and \ref{sec:failedsnr}, but the overall contribution from the helium-shell production of these isotopes, compared to full r-process production sources, is small \citep{Cote.ea:2021}. This might imply that, while the vast majority of these nuclei that have existed in the history of the Galaxy came from r-process nucleosynthesis, some of these nuclei present in the early solar nebula may instead have been due to explosive helium burning in a massive star. This would further suggest that the last contributing r-process event occurred simply too long before the solar system formation, or too far away, to have meaningfully contributed. Further detailed discussions of isotopic features associated with such neutron bursts in the shock wave of CCSNe have been given in \citep{Liu.Lugaro.ea:2024} and \cite{Battino.ea:2024}. Fig.3 of the latter reference shows, however, how difficult it is to obtain similar productions in comparison to the solar system values for the three isotopes $^{107}$Pd, $^{129}$I, and $^{182}$Hf, discussed above. 

Recent observations by the Rosetta mission with respect to Xe isotopes \citep{Cassata:2025,Cassata.ea:2026} seem to indicate that two different r-process sources contributed. One of them possibly mixed with a p-process component, which could possibly be explained by a $\nu$r-process, and may take place in neutron-rich ejecta experiencing an intensive neutrino flux \citep{Xiong.Martinez.ea:2024}.

\section{Conclusions}\label{sec13}

We have learned much in almost 70 years of studying the r-process \citep[see e.g.][]{Burbidge.Burbidge.ea:1957, seeger.fowler.clayton:1965,Kodama.Takahashi:1975, Hillebrandt:1978,Cowan.Thielemann.Truran:1991,Kratz.Bitouzet.ea:1993,Woosley.Wilson.ea:1994,Arnould.Goriely:2001,Arnould.Goriely.Takahashi:2007,Thielemann.Arcones.ea:2011,Kajino.Aoki.ea:2019,Cowan.Sneden.ea:2021,Hotokezaka.Wanajo:2026}, but there are many remaining questions. Observational and theoretical results suggest that some type of compact binary merger is responsible for the main, or the majority, of r-process production over the history of, and throughout, the Galaxy. Nevertheless, multiple astrophysical sites have been proposed that contribute to the production of these isotopes and elements, and observational indications exist that such additional sites play a role as well. New nuclear physics experiments, in particular at a variety of radioactive ion beam facilities, like FRIB, RIKEN, CERN/ISOLDE, and FAIR, along with continued high-resolution spectroscopic studies of metal-poor Galactic halo stars and further observational constraints on occurrence frequencies of the suggested astrophysical sites, as well as detailed nucleosynthesis models will be needed to make further progress in understanding how all these neutron-rich isotopes are formed in nature. 

While very early proposals thought of making heavy r-process elements already in the Big Bang, starting essentially with neutrons \citep{Alpher.Bethe.Gamow:1948}, an idea that was rejuvenated on several occasions \citep{Applegate:1988,Kajino.Mathews.Fuller:1990,rauscher1994,Roepke.Blaschke.ea:2025}, present-day observations of extremely and very metal-poor (EMP and VMP) stars show abundance patterns starting at the lowest metallicities with typical core-collapse supernova ejecta compositions. This is followed by stars showing limited (weak) r-process abundances and finally r-process-rich main and robust r-process abundances (r-I and r-II). This is a clear sign that - like all other elements heavier than H, He, and Li - the r-process elements also originate from stellar explosions. The question is which sites can be identified. 

If the solar r-process abundance pattern would be due to one dominant site, the basic requirements for such a site are presented in Fig.\ref{fig:rosswog17} in terms of necessary ejecta amounts vs. the occurrence frequency in the Galaxy, completely independent of the type of astrophysical site. If, on the other hand, we assume neutron star mergers are the dominant site, a frequency of the order of 0.3\% of the CCSNe rate is required \citep{Chen.Li.ea:2024}, corresponding to a rate of $R=$500-600Gpc$^{-3}$y$^{-1}$, in comparison to the limits given by \cite{Abbott.eaX:2021} of 310$^{+490}_{-240}$Gpc$^{-3}$y$^{-1}$ and also the recently published large uncertainty range of 7.6 - 250 Gpc$^{-3}$yr$^{-1}$ from the few identifications based on gravitational wave observations by \cite{Abac.ea:2025}, \citep[see also the discussion in][]{Fishbach.ea:2026}. In addition, we know from observations of low-metallicity limited-r stars that at least two different sites are needed to produce r-process nuclei. Such limited-r stars are first seen at metallicities of [Fe/H]$<$-4 (see Fig.\ref{fig:farouqFeH}).

This would be consistent with events taking place with a frequency of 1 out of 10 CCSNe (seen first below a metallicity of [Fe/H]$<$-5), in agreement  with the expected frequency of magneto-rotational supernovae that leave neutron stars with high magnetic fields (magnetars) as central remnants. An additional aspect is that \cite{Farouqi.ea:2022} and \cite{Farouqi.ea:2025} find a correlation of Eu with Fe in these limited-r stars, hinting at a co-production of Fe and Eu in the originating source and also pointing to the origin from magneto-rotational supernovae. For heavier elements \cite{Farouqi.ea:2025} (see their Table 1) find no correlation with Fe, indicating no or negligible co-production of Fe (in comparison to solar ratios) in the corresponding sites, which is consistent with the predicted abundances in their ejecta. In this review we have discussed neutron star mergers as well as collapsars/hypernovae and neutron star - black hole mergers as origin for the heavy r-process elements. \cite{Farouqi.ea:2025} find differences in the Pearson and Spearman correlation coefficients for r-I and r-II stars, which could point to different sources (possibly the two above mentioned sites, where the latter ones show an up to a factor of 10 higher r-process production and might therefore be good candidates for r-II stars). However, we should be aware, that with an increasing number of regular CCSNe contributing to an Fe floor for these rare events, such a correlation analysis between Fe and r-process elements becomes less meaningful. The observations of \cite{Roederer.eauniversal:2022} and \cite{Racca.ea:2025} indicate a universality of the r-process among heavy elements in r-I and r-II stars, arguing for similar production conditions for the heavy r-process elements, while permitting for a different total r-process production. This universality seems to be broken in the cases of actinide boosts among r-II stars \citep{Mashonkina:2014,Shah.ea:2026}, possibly underlining the neutron star - black hole merger origin \citep{Wanajo.Fujibayashi.ea:2024}.
Thus, despite many advances in experiments, theory and observations, there remain a number of open questions for the future. 

\backmatter

\begin{figure}[h]
\centering
\includegraphics[width=0.4\textwidth]{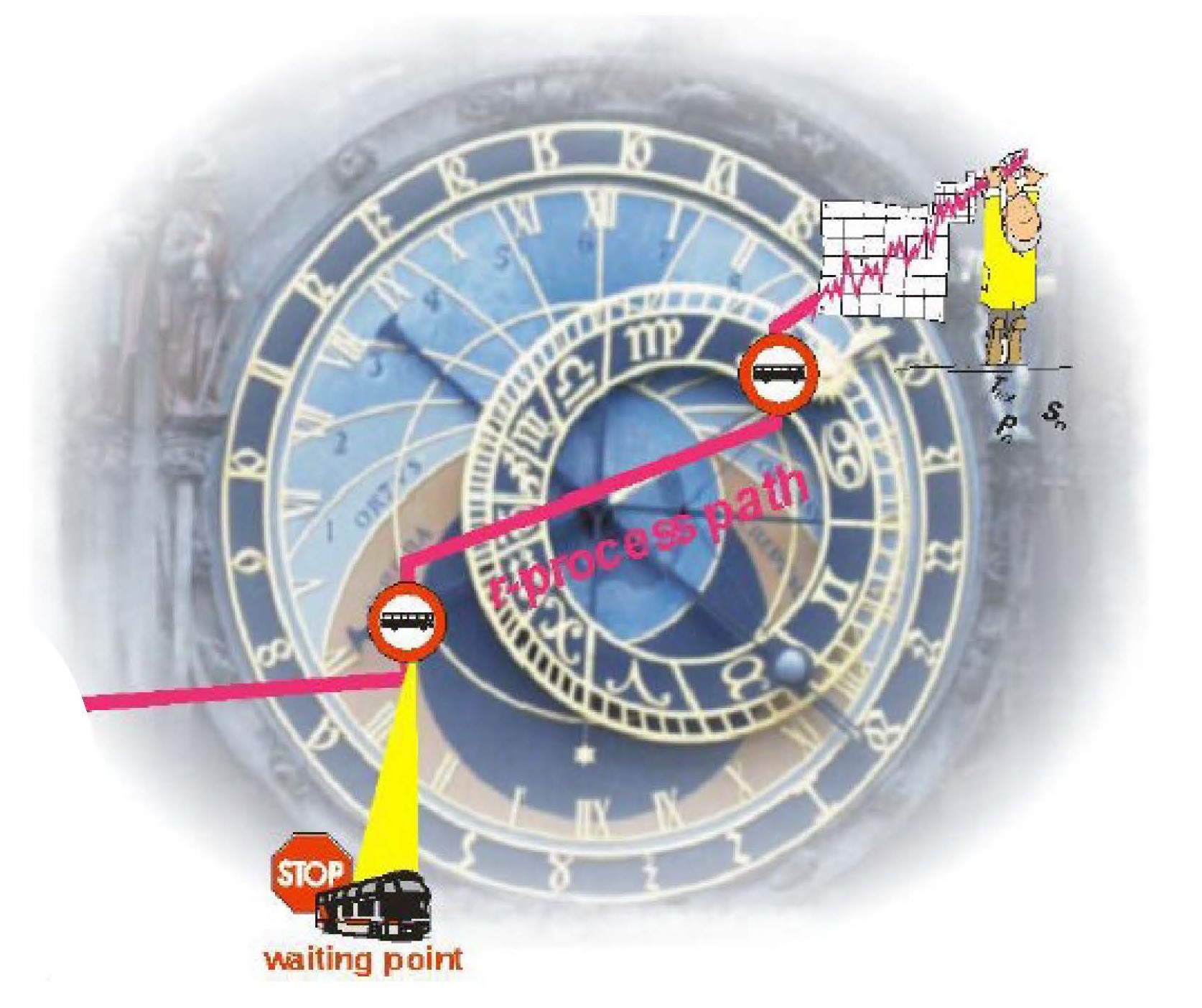}
\includegraphics[width=0.55\textwidth]{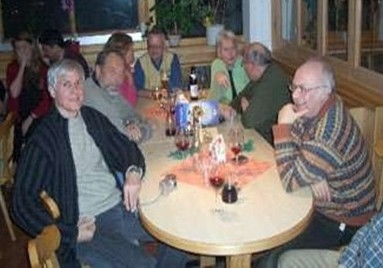}
\caption{Left: One of Karl-Ludwig Kratz' famous plots about the r-process path (with the astronomical clock of the historical Prague City Hall in the background), right: K.-L. Kratz (center) jointly with the authors (front) and a group of participants at one of the famous Russbach meetings, a conference series which he started.}\label{PraguepathRussbach}
\end{figure}

\bmhead{Acknowledgements}


This EPJA volume is dedicated to our late friend and colleague Roberto Gallino, whose pioneering and pivotal efforts for understanding the s-process have been addressed in Sections \ref{sec1} and \ref{sec2}. Initially the current authors were invited to contribute a review about the r-process to this memorial volume jointly with our late friend and colleague Karl-Ludwig Kratz, who unfortunately passed away only a month after that invitation \citep[see Fig.\ref{PraguepathRussbach} and][]{Aprahamian.ea:2025}.
We therefore also want to dedicate this paper to Karl-Ludwig (a  winner of the Glen Seaborg Award of the ACS and the Hans A. Bethe Prize of the APS), with whom we authored many publications related to site-independent studies of the r-process and Galactic cosmochronology. We insert here a picture taken from years ago at a scientific meeting. We learned much from both of these men, Roberto and Karl-Ludwig, both in life and in science. We will miss them dearly, but will continue to pursue scientific studies that we shared during their lives.

We thank an excellent referee who combined a very positive overall review with many detailed comments for further improvements. 
With respect to preparing the present review we want to thank Almudena Arcones, Andreas Bauswein, Khalil Farouqi, Anna Frebel, Ore Gottlieb, Aldana Grichener, Terese Hansen, Agnieszka Janiuk, Alexander Ji, Oliver Just, Nan Liu, Maria Lugaro, Gabriel Martinez-Pinedo, Brian Metzger, Ken Nomoto, YongZhong Qian, Ian Roederer, Stephan Rosswog, Shivani Shah, Noam Soker, Chris Sneden, Rebecca Surman, Shinya Wanajo, Tony Wallner, and Zewei Xiong for valuable discussions and advice, aided by communications within the r-Process Alliance and the International Research Network for Nuclear Astrophysics IReNA (NSF grants OISE 1927130 and PHY 14-30152).

\end{document}